\documentclass[usenatbib]{mn2e}
\usepackage{psfig}

\def\gsim{~\rlap{$>$}{\lower 1.0ex\hbox{$\sim$}}}
\def\lsim{~\rlap{$<$}{\lower 1.0ex\hbox{$\sim$}}}
\def\d{{\rm d}}
\def\c{{\rm c}}

\begin{document}

\title{Statistics of Sunyaev-Zel'dovich Cluster Surveys} 
\author[Andrew~J.~Benson, Christian~Reichardt \& Marc~Kamionkowski]{Andrew~J.~Benson, Christian~Reichardt \& Marc~Kamionkowski \\
\noindent California Institute of Technology, MC 130-33, Pasadena, CA 91125, U.S.A.}

\maketitle

\begin{abstract}
We describe a detailed analytic model for predicting statistical
quantities (such as number counts, redshift distributions and sizes)
of clusters detected in blank-field, thermal Sunyaev-Zel'dovich effect
experiments. We include in this model the possibility of non-Gaussian
density perturbations in the early Universe and also describe a simple
model for the effects of preheating on cluster Sunyaev-Zel'dovich
effect fluxes. We use this model to explore the current state of the
theoretical uncertainties present in this type of analytic modelling,
highlighting where further improvement will be necessary to fully
exploit forthcoming surveys. We then go on to explore the constraints
on cosmological parameters, the presence of any non-Gaussianity and
the degree of cluster preheating which may be obtained from both the
{\sc bolocam} and {\sc planck} experiments. We find that, providing
redshifts can be measured for all detected clusters, the {\sc bolocam}
experiment may provide detections of non-Gaussianity or preheating and
could give approximate measurements of these effects if prior
knowledge of the various cosmological parameters is taken into
account. The {\sc planck} experiment Sunyaev-Zel'dovich effect cluster
survey is predicted to provide highly accurate ($\sim 5\%$)
measurements of the degree of non-Gaussianity and preheating while
also providing measurements of several cosmological parameters to
accuracies of a few percent independent from those constraints that
will be derived from its detections of primordial cosmic microwave
background anisotropies.
\end{abstract}


\section{Introduction}

The Sunyaev-Zeldovich (SZ) effect (\citealt{sz72}; for recent reviews
see \citealt{reph95} and \citealt{birk99}), the distortion of the
cosmic microwave background (CMB) spectrum due to the scattering of
photons from electrons in hot, ionized gas (typically in clusters), is
about to come of age as a cosmological probe. To date, the SZ effect
has been mapped in a few clusters through targeted observations
\citep{carlstrom00,joy01}, but planned experiments such as {\sc
AmiBA}, {\sc acbar}, {\sc bolocam}, {\sc cbi}, {\sc dasi} and {\sc
planck} will, for the first time, carry out statistically useful
surveys of clusters in the SZ effect. Since the formation rate of
clusters is strongly dependent on the values of cosmological
parameters we may hope to place constraints on these parameters
through such SZ surveys
\citep{thomas89,scar93,barbosa96,ecf96,dasilva00,refregier00,majumdar00,holder00,dasilva00b,fan01,xuewu01,seljak00,springel01,gendin01,kay01}. Furthermore,
as the SZ effect probes the baryonic content of clusters it has the
potential to reveal the effects, if any, of physical processes such as
feedback and preheating on the cluster gas
\citep{majumnath00,springel01,majum01}.

For these reasons several studies of the SZ effect have been made
recently, using both numerical methods (e.g. \citealt{dasilva00}),
analytic techniques (e.g. \citealt{holder01,delabrouille01}) and
combinations of both (e.g. \citealt{kay01}). Numerical methods have
the advantage of giving a full treatment of the complex gas physics
that determines the strength of the SZ signal from each cluster, but
suffer from being computationally expensive. Analytic methods on the
other hand are generally highly computationally efficient, but can
only treat the gas dynamics in an approximate way. The ideal solution
would be to build a fast and accurate analytic model tuned to
reproduce the results of numerical simulations. As the first
blank-field SZ surveys are expected to yield results in the next year,
the time is right to explore in more detail the current theoretical
uncertainties in analytic SZ models in order to highlight where
further improvement is necessary. It will also be crucial to ascertain
just how well proposed experiments can constrain both cosmological
parameters and the distribution of gas within clusters. In this work
we will address both of these questions.

The remainder of this paper is arranged as follows. In
\S\ref{sec:model} we describe our calculations of the thermal SZ
fluxes of clusters. In \S\ref{sec:results} we contrast predictions for
two proposed experiments, explore some of the key theoretical
uncertainties in computing SZ survey statistics and derive the
constraints obtainable on cosmological and gas distribution parameters
in the near future. Finally, in \S\ref{sec:disc} we give our
conclusions.

\section{Model}
\label{sec:model}

We begin by describing how we compute the observable properties of SZ
clusters. Our model determines only the thermal SZ signal, since we do
not predict the peculiar velocities of clusters needed to calculate
the kinetic signal. This is not a significant problem since the
magnitude of the kinetic effect is over one order of magnitude smaller
than that of the thermal effect except at frequencies close to the
null-point in the thermal effect spectral profile \citep{kay01}. In
\S\ref{sec:halos} we describe the calculation of the dark-matter halo
properties and in \S\ref{sec:bary} we discuss how these halos are
filled with baryonic material. Finally, in \S\ref{sec:SZ} we detail
how the observable SZ properties are determined from this gas
distribution.

\subsection{Halo Distribution}
\label{sec:halos}

We assume that dark-matter halos in the Universe can be characterised
by three quantities, their mass, $M$, the redshift at which they
formed, $z_{\rm f}$, and the redshift at which they are observed,
$z_{\rm o}$. From these quantities we can determine, for given
cosmological parameters, the virial radius and temperature of the halo
from the spherical top-hat collapse model (these properties are
assumed to be fixed at the formation redshift and to remain fixed
until the cluster is observed). To calculate the abundance of clusters
as a function of these three parameters we use the Press-Schechter
theory \citep{ps74,bcek91,bower91}, plus extensions described by
\citet{lc93} and \citet{sasaki94} that allow the distribution of
formation redshifts to be determined.

The Press-Schechter theory predicts that the number of clusters of
mass $M$ to $M+{\rm d}M$ per unit comoving volume, $V$, observed at
redshift $z_{\rm o}$ is
\begin{equation}
{{\rm d}^2 n \over {\rm d}M {\rm d} V} = {f \rho_{\rm crit}\Omega_0 \over M} {\partial y \over \partial M} P(y),
\end{equation}
where $\rho_{\rm crit}$ is the critical density of the Universe at the
present day, $y=\delta_{\rm c}(z)/\sigma(M)$, with $\delta_{\rm c}(z)$
being the critical linear overdensity for collapse at redshift $z$
divided by the growth factor (which is normalised to unity at the
present day), $\sigma(M)$ being the linear theory mass variance at the
present day in a sphere containing mass $M$ on average, $P(y)$ being
the probability distribution function for density perturbations in the
early universe and $f = 1/\int_0^\infty P(y) {\rm d}y$. Typically,
$P(y)$ is taken to be a Gaussian (as expected from traditional models
of inflation), but we will also consider other forms in this
paper. This distribution is easily expressed in terms of $z_{\rm o}$
using the volume-redshift relation for Friedmann cosmologies
\begin{equation}
{\d V \over \d z_{\rm o}} = 4 \pi r^2(z_{\rm o})\c H^{-1}(z_{\rm o}),
\end{equation}
where $c$ is the speed of light, $r(z)$ is the coordinate distance to
redshift $z$, $H(z)$ is the Hubble constant at that redshift and we
have computed the volume over the whole sky. Therefore
\begin{equation}
{{\rm d}^2 n \over {\rm d}M {\rm d} z_{\rm o}} = {{\rm d}^2 n \over {\rm d}M {\rm d} V} {{\rm d} V \over {\rm d} z_{\rm o}} .
\end{equation}
Finally, we consider two models for the distribution of halo formation
redshifts. In the model of \citet{sasaki94} the fraction of clusters
of mass $M$ seen at redshift $z_{\rm o}$ which formed between
redshifts $z_{\rm f}$ and $z_{\rm f} + {\rm d}z_{\rm f}$ is given by
\citep{verde01}
\begin{equation}
{{\rm d} f \over {\rm d} z_{\rm f}} = {1 \over P(y_{\rm o})} {{\rm d} P(y_{\rm f}) \over {\rm d} y} {\partial y \over \partial z_{\rm f}}.
\end{equation}
Therefore, the number of clusters seen in the range $M$ to $M+{\rm
d}M$, $z_{\rm f}$ to $z_{\rm f} + {\rm d}z_{\rm o}$ and $z_{\rm o}$ to
$z_{\rm f} + {\rm d}z_{\rm f}$ is
\begin{equation}
{{\rm d}^3 n \over {\rm d}M {\rm d} z_{\rm o} {\rm d}z_{\rm f}} = {{\rm d}^2 n \over {\rm d}M {\rm d} V} {{\rm d} V \over {\rm d} z_{\rm o}} {{\rm d} f \over {\rm d} z_{\rm f}}.
\end{equation}

\begin{figure}
\psfig{file=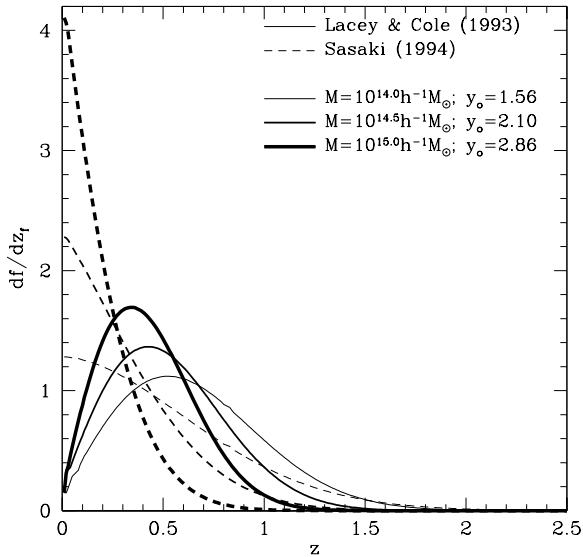,width=80mm}
\caption{Comparison of formation redshift distributions. The curves
show the fractional formation rate as a function of redshift for a
clusters of masses $10^{14}$, $10^{14.5}$ and $10^{15}h^{-1}M_\odot$
(thin, heavy and very heavy lines respectively) observed at $z_{\rm
o}=0$ in a $\Lambda$CDM cosmology. The corresponding value of $y_{\rm
o}$ is given for each cluster in the figure. The solid lines indicate
the \protect\citet{lc93} calculation while the dashed lines indicate
the \protect\citet{sasaki94} calculation.}
\label{fig:formcompare}
\end{figure}

An alternative derivation of the distribution of halo formation times
due to \citet{lc93} gives the result
\begin{eqnarray}
{{\rm d} f \over {\rm d}z_{\rm f}} & = & {{\rm d} \over {\rm d}z_{\rm f}} \int_{M/2}^M \left( {2 \over \pi} \right)^{1/2} {M\over M^\prime} {y_{\rm f}-y_{\rm o} \over (\sigma(M^\prime)^2/\sigma(M)^2-1)^{3/2}} \nonumber \\ 
 & & \times {\sigma(M)^2 \over \sigma(M^\prime)^2} {{\rm d}\ln \sigma \over {\rm d}\ln M} \nonumber \\
 & & \times \exp \left[ - {(y_{\rm f}-y_{\rm o})^2/2 \over \sigma(M^\prime)^2/\sigma(M)^2-1} \right] {\rm d} M^\prime ,
\end{eqnarray}
for the case of Gaussian initial conditions. The difference between
the two results is due to the different way in which ``formation'' is
defined in the two approaches. \citet{lc93} define the formation event
as that time at which a halo first had a progenitor at least half as
massive as itself. \citet{sasaki94} instead derives expressions for
the formation and destruction rates of halos under the assumption that
the destruction rate has no preferred mass
scale. Figure~\ref{fig:formcompare} compares the two
formation-redshift distributions for clusters of masses $10^{14}$,
$10^{14.5}$ and $10^{15}h^{-1}M_\odot$ (thin, heavy and very-heavy
lines respectively) observed at $z_{\rm o}=0$ in a $\Lambda$CDM
cosmological model. Note that although the shapes of the two
distributions differ, both decline rapidly at high redshifts and the
mean redshift of formation becomes lower as the cluster mass (and
hence $y_{\rm o}$) increases.

It is apparent that the two distributions are rather different, with
the \citet{lc93} model predicting a much higher mean redshift of
formation. The question of which is the best distribution is almost
certainly problem dependent. \citet{lc94} show that their calculation
accurately describes the results of N-body simulations given their
definition of formation time. Note that the expression derived by
\citet{sasaki94} is qualitatively similar to that derived by
\citet{lc93} using the mass-halving times for single trajectories,
although the \citet{lc93} expression contains a dependence on the
shape of the dark-matter power spectrum whereas that of
\citet{sasaki94} does not. However, for our work we are really
interested in knowing the properties of gas in a halo. Halos of given
mass that formed earlier will be denser and hotter. In reality, the
process of cluster formation is a continuous one and so it is not
clear just how the formation redshift should affect the gas
properties. In this work we will therefore consider both
distributions, and also the case of $z_{\rm f}=z_{\rm o}$.  Two of
these possibilities are limiting cases. The case $z_{\rm f} = z_{\rm
o}$ is clearly the lower limit on formation redshift, while the
distribution proposed by \citet{lc93} should give a reasonable upper
limit as it is difficult to see how the observed gas properties could
be determined by the physical state at any earlier time (i.e. when
less than half of the final mass of the cluster is in place). Our
default distribution will be the Sasaki model.

With $P(y)$ a Gaussian the Press-Schechter theory predictions differ
significantly from the results of numerical simulations of structure
formation (by a factor close to 2 at the characteristic mass
$M_*$). Several fitting formulae have been proposed which produce a
much better agreement with numerical results. We will consider two
such fitting formulae, those proposed by \citet{st99} and
\citet{jenkins00} (hereafter ST and J2000 respectively), which can be
described by
\begin{eqnarray}
P_{\rm ST}(y) & = & 0.3223 \sqrt{2 a \over \pi} \left[1+{1 \over (a y^2)^q}\right] \exp\left(-{ a y^2\over 2}\right) \\
P_{\rm J2000}(y) & = & {A \over y} \exp\left(-\left|\ln\left[{y\over 1.686}\right]+c\right|^p\right),
\end{eqnarray}
where $a=0.707$ and $q=0.3$ for the ST formula and $A=0.301(307)$,
$c=0.64(0.61)$ and $p=3.88(3.82)$ for the J2000 formula in the
$\Lambda$CDM($\tau$CDM) cosmology considered in \S\ref{sec:results}.

To explore non-Gaussian initial conditions we make use of the
log-normal probability distribution proposed by \citet{rgs},
\begin{equation}
P_{\rm RGS}(y) = {f C \over \sqrt{2 \pi} |A|} \exp\left[-{x^2(y)\over 2}-|A|x(y) \right],
\end{equation}
where $x(y)=\ln(B+C y |A|/A)/|A|$, with $A$, $B$ and $C$ as defined by
\citet{rgs}. The degree of non-Gaussianity is fully specified by the
parameter $A$ (with $A=0$ corresponding to the Gaussian limit). We
will, however, characterize non-Gaussian models by the more intuitive
variable $G$, defined by \citet{rgs} as the number of $>3\sigma$ peaks
in the non-Gaussian distribution relative to the number in a Gaussian
distribution (such that $G=1$ represents the Gaussian limit).

\subsubsection{Cosmic Variance}
\label{sec:cosvar}

The above calculations allow us to calculate the mean abundance of
dark-matter halos as a function of their mass, and redshifts of
observation and formation for any given cosmological
parameters. However, for SZ surveys covering relatively small fields
of view it is important to assess the effects of sample variance
(a.k.a. cosmic variance) in order to accurately determine the ability
of the survey to discriminate between models. In fact, the sample
variance is expected to be rather small (close to Poissonian) since
the SZ effect probes a wide range of redshifts such that any intrinsic
correlations between clusters are diluted (as can be estimated using
analytical calculations of the cluster bias to compute their angular
correlation function). To measure the sample variance we make use of
the Hubble Volume simulations which were carried out by the VIRGO
Consortium and which are publically available \citep{evrard98}. These
large N-body simulations have been used to construct catalogues of
dark-matter halos (listing the mass and observed redshift) as seen
along a past lightcone.

To compute the cosmic variance from these simulations we construct
mock surveys of the required angular size by choosing a random line of
sight through the simulation lightcone and assigning each cluster in
the field of view the SZ flux computed from our model for a cluster
of the same mass and observed redshift and with a formation redshift
drawn at random from the distributions described above (note that this
assumes that the spatial distribution of halos of given mass and
observed redshift is independent of their formation redshift, which is
unlikely to be correct in detail). We then compute the statistic of
interest from this mock survey. Repeating for several mock surveys we
obtain a measure of the variance in the statistic. Results from these
calculations will be shown in \S\ref{sec:rescosmo}.

\subsection{Baryon Distribution}
\label{sec:bary}

\subsubsection{No Preheating/No Cooling}
\label{sec:nopre}

Numerical simulations of gas in CDM clusters, for example the Santa
Barbara cluster comparison project (SBCCP) of \citet{frenk99}, show
that, in the absence of heating and cooling processes (other than
shocks and adiabatic compression) the baryonic material of a cluster
has a density profile well described by a beta-model \citep{cav76},
and is close to hydrostatic equilibrium, with a temperature at the
virial radius equal to approximately half the virial temperature of
the cluster. For each cluster in our calculations we determine the
virial temperature from the spherical top-hat collapse model and the
total mass of gas using $M_{\rm gas} = (\Omega_{\rm
b}/\Omega_0)M$. This gas is distributed in the cluster as a beta-model
with $\beta = 0.9$ and a core radius equal to 0.09 of the virial
radius. We set the gas temperature at the virial radius to one half of
the virial temperature and determine the temperature at smaller radii
by assuming hydrostatic equilibrium. As expected, this reproduces well
the profile of the SBCCP cluster. Beyond the virial radius we
extrapolate the pressure and density as power-laws which also provides
a good fit to the SBCCP cluster (hydrostatic equilibrium is not a good
assumption here as infall is occurring).

\subsubsection{Preheated and/or Cooled Profiles}
\label{sec:descpre}

It is well known that, in the absence of radiative cooling or any
other process (other than infall-driven shocks) the relation between
the X-ray luminosity and temperature of clusters differs significantly
from that observed. One possible solution is that the intra-cluster
medium (ICM) gas was preheated before it collapsed into the cluster,
increasing its entropy and resulting in a shallower density
profile. (For further discussion of these points see, for example,
\citealt{tozzi01}.)

In the case of no preheating the entropy of the cluster gas arises
almost entirely from the shock it experiences as it accretes onto the
cluster. If we assume that the gas is in hydrostatic equilibrium, then
its ``entropy'' density as a function of radial distance $r$ from the
cluster centre is given by
\begin{eqnarray}
K_{\rm shock}(r) & = & {1 \over \rho_{\rm g}^{\gamma}(r)}\left[ - \int_{r_{\rm vir}}^r {{\rm G}M(r^\prime)\rho_{\rm g}(r^\prime) \over r^{\prime 2}}{\rm d}r^\prime \right. \nonumber \\
 & & \left. + {\rho_{\rm g}(r_{\rm vir}) \over \mu {\rm m_H}} {\rm k_B}T_{\rm g}(r_{\rm vir})\right],
\end{eqnarray}
where $\rho_{\rm g}$ and $T_{\rm g}$ are the gas density and
temperature respectively, $\gamma=5/3$ is the ratio of specific heats
for the gas, $M(r)$ is the cluster mass profile, $r_{\rm vir}$ the
cluster virial radius and $\mu=0.59$ the mean atomic mass appropriate
to a fully ionized primordial gas (with hydrogen mass fraction
$X=0.76$).

\begin{table*}
\caption{Parameters of the two experiments, {\sc bolocam} and {\sc planck}, considered in this work.}
\label{tb:expparams}
\begin{center}
\begin{tabular}{cccccc}
Experiment & Solid angle (sq. deg.) & Beam FWHM & Central frequency & Frequency response & Background, $y_{\rm bg}+3\sigma_{\rm bg}$ \\
\hline
{\sc bolocam} & 1 & $1^{\prime}$ & 143 GHz & Top-hat, $\Delta\nu/\nu=0.07$ & $1.8\times 10^{-5}$ \\
{\sc planck} & 41253 (all sky) & $8^\prime$ & 143 GHz & Gaussian, $\sigma=22.5$GHz & $1.0\times 10^{-5}$
\end{tabular}
\end{center}
\end{table*}

Motivated by X-ray studies \citep{ponman99} we introduce a minimum
entropy such that the entropy profile becomes,
\begin{equation}
K(r) = \left\{ \begin{array}{ll} K_{\rm min} & r < r_{\rm crit} \\ K_{\rm shock}(r) & r \geq r_{\rm crit} \end{array} \right. ,
\end{equation}
where $r_{\rm crit}$ is defined by $K_{\rm shock}(r_{\rm crit})=K_{\rm
min}$. Assuming hydrostatic equilibrium and an NFW density profile for
both dark matter and shock-heated gas the corresponding density
profile of the preheated gas can be obtained (by expressing the
equation of hydrostatic equilibrium in terms of $\rho_{\rm g}$ and
$K$, and then solving for $\rho_{\rm g}$) and is given by (S.-P. Oh,
private communication)
\begin{equation}
\rho_{\rm g}(r) = \left\{ \begin{array}{ll} \left[\left({\Omega_{\rm b}\over \Omega_0} \rho_{\rm NFW}(r)\right)^{\gamma -1} - {\gamma\over \gamma-1}{{\rm G} M_{\rm h}\over r_{\rm s}} \right. \\ \times \left(\ln (1+c_{\rm NFW}) - {c_{\rm NFW}\over(1+c_{\rm NFW})}\right)^{-1} \\ \left. \times {1\over K_{\rm min}}\left(  {\ln(1+x_{\rm crit})\over x_{\rm crit}} - {\ln(1+x)\over x} \right) \right] ^{{1\over 1-\gamma}} & r < r_{\rm crit} \\ {\Omega_{\rm b}\over \Omega_0} \rho_{\rm NFW}(r) & r \geq r_{\rm crit} \end{array} \right.
\end{equation}
where $c_{\rm NFW}$ is the concentration parameter of the NFW profile
(defined as the ratio of virial to scale radii and computed using the
method given by \citealt{nfw97}). From this the corresponding pressure
and temperature profiles are easily found. The effect of the minimum
entropy is to reduce the central density and pressure of gas in halos,
the effect being strongest in lower-mass halos. This necessarily
reduces the SZ flux of the halos and so affects the abundances and
redshift distributions of SZ selected samples. We choose to express
the minimum entropy density as $K_{\rm min}=10^{34} K_{34}
h^{-4/3}$ergs cm$^{-2}$ g$^{-5/3}$. \citet{tn00} found that values of
$K_{34}\approx 0.12$ and $0.25$ were required to explain the
properties of groups and clusters with X-ray temperatures below and
above 2\,keV respectively, or alternatively that a redshift dependent
value of $K_{34}=0.5/(1+z)$ fitted the data over the whole temperature
range for which data was available\footnote{Note that
\protect\citet{tn00} define $K_{\rm min}=10^{34} K_{34}$\,ergs
cm$^{-2}$ g$^{-5/3}$, i.e. without the explicit $h$ dependence of our
definition. We have adjusted their values of $K_{34}$ accordingly.}.

There remains some doubt as to whether preheating is necessary at
all. \citet{dave01} use hydrodynamical simulations which include
radiative cooling to show that a good fit to the cluster X-ray
luminosity-temperature relation can be achieved when cooling is
included without the need for any preheating (see also
\citealt{pearce00,bryan00,muanwong01}).

Unfortunately, the effects of cooling on cluster gas profiles are more
difficult to ascertain analytically, and this will be a key area in
which analytic modelling must improve to fully exploit future SZ
surveys to constrain the physical processes affecting cluster gas. The
high-density/low-entropy inner regions of the cluster should cool in
times much shorter than the Hubble time. Current evidence from
numerical simulations (e.g. \citealt{pearce00,muanwong01}) suggests
that as gas cools the surrounding higher entropy gas flows inwards to
take its place. If this is indeed the case the cooling will also
introduce a minimum entropy into cluster gas profiles. As such, we
expect our model of preheating to produce qualitatively similar hot
gas profiles to those resulting from radiative cooling. Therefore,
lacking at present any better way to parameterise the effects of
cooling we will consider preheated models only, and recognise that
what looks like preheating may in fact be the result of radiative
cooling.

\subsection{SZ Calculation}
\label{sec:SZ}

Our calculation of the SZ flux essentially follows that of
\citet{kay01}. Having determined the density and temperature profiles
of the electrons in each cluster we determine the Compton parameter as
a function of angular radius, $y(\theta)$, using
\begin{equation}
y(\theta) = {k_{\rm B} \sigma_{\rm T} \over m_{\rm e}{\rm c}^2} \int _{-\infty}^{\infty} n_{\rm e} T_{\rm e} {\rm d}l,
\end{equation}
where $k_{\rm B}$ is Boltzmann's constant, $\sigma_{\rm T}$ is the
Thomson cross section, $m_{\rm e}$ is the mass of an electron and the
integral is taken along the line of sight passing a distance $\theta$
away from the centre of the cluster. The electron density is $n_{\rm
e}=0.88 \rho_{\rm g}/m_{\rm H}$, where $m_{\rm H}$ is the mass of a
hydrogen atom, as appropriate for a fully ionized primordial gas (with
hydrogen mass fraction $X=0.76$). Although the above integral extends
from $-\infty$ to $+\infty$, in practice it converges with limits of
order the virial radius of the halo.  This profile is then convolved
with a beam to mimic instrumental effects, giving
\begin{equation}
y_{\rm s}(\theta) = \int y(\hbox{\boldmath{$\theta$}} ^\prime) b(|\hbox{\boldmath{$\theta$}}-\hbox{\boldmath{$\theta$}}^\prime|) {\rm d}^2\hbox{\boldmath{$\theta$}}^\prime ,
\end{equation}
where $b(\theta)$ is the beam profile, which we take to be a Gaussian.
The total SZ flux increment or decrement at frequency $\nu$ is then
found using
\begin{equation}
S_{\nu} = 2 \pi S_0 G(x_0) \int _0^{\theta _{\rm b}} y_{\rm s}(\theta) \theta {\rm d} \theta ,
\label{eq:intflux}
\end{equation}
where $S_0=2.29 \times 10^4$ mJy arcmin$^{-2}$ with $\theta$ in arcmin
and $\theta_{\rm b}$ is the radius at which the Compton $y$ parameter
of the observed cluster profile falls below the background. We define
\begin{equation}
G(x_0) = \int_0^\infty g(x) f(x-x_0) {\rm d}x,
\end{equation}
where $x = {\rm h}\nu / {\rm k_B} T_{\rm CMB}$, $h$ is Planck's
constant, $T_{\rm CMB}=2.725$K is the mean temperature of the CMB
\citep{mather99},
\begin{equation}
g(x) = {x^4 {\rm e}^x \over ({\rm e}^x - 1)^2} \left[ x {{\rm e}^x + 1 \over {\rm e}^x -1} - 4 \right]
\end{equation}
is the frequency dependence of the thermal SZ effect and $f(x)$ is the
filter response function, which we will take to be either a Gaussian
or a top-hat.

\begin{figure*}
\begin{tabular}{c@{}c@{}c}
\psfig{file=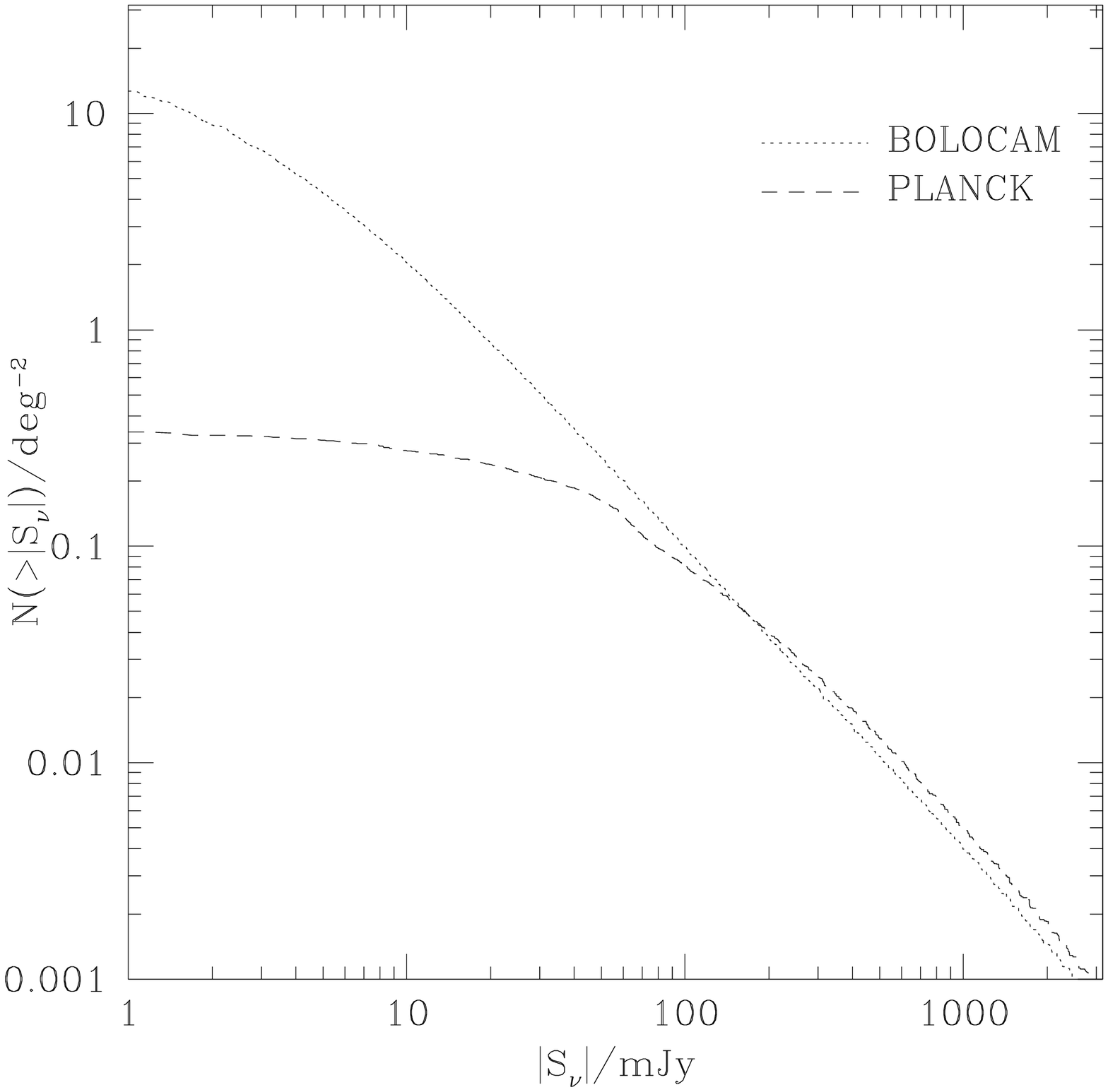,width=58mm} &
\psfig{file=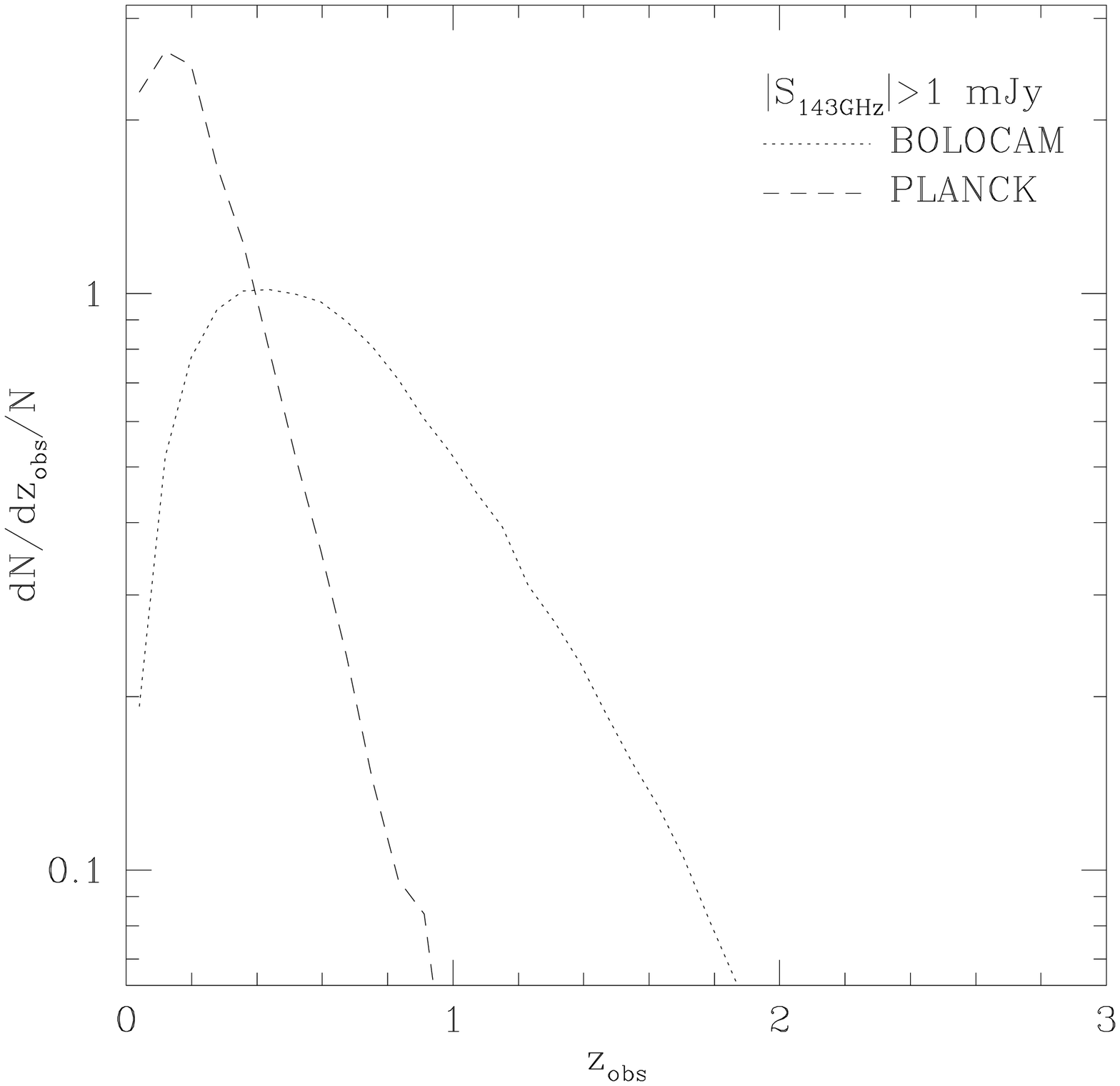,width=58mm} &
\psfig{file=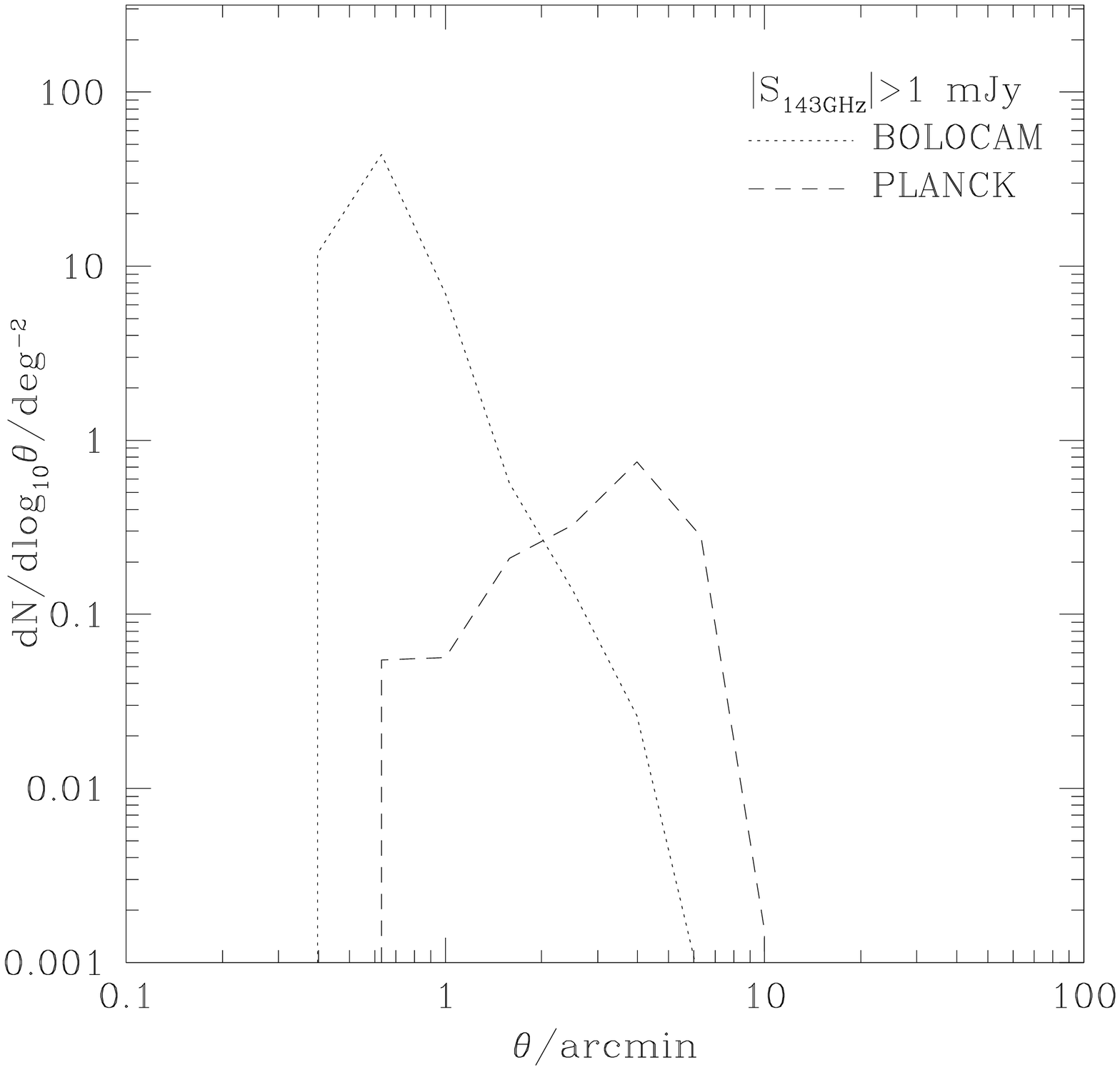,width=58mm}
\end{tabular}
\caption{Properties of SZ clusters as detected by {\sc bolocam}
(dotted lines) and {\sc planck} (dashed lines). The left-hand panel
shows counts as a function of flux at 143GHz, the centre panel shows
the redshift distribution of clusters brighter than $|S_{143{\rm
GHz}}|=1$mJy, and the right hand panel shows the distribution of
angular sizes for the same clusters. All results are for the
$\Lambda$CDM cosmology with the J2000 $P(y)$.}
\label{fig:BvP}
\end{figure*}

\subsubsection{Background}

The limiting flux for SZ detections will depend upon the SZ background
due to the summed contributions of all unresolved sources. This
background can be characterized by a mean Compton $y$ parameter and an
rms fluctuation about this value. We can calculate the background and
its fluctuations from our model using an approach similar to that
described by \citet{bartelmann00} (similar calculations have been used
to compute the power spectrum of the SZ effect by
\citealt{bruscoli00,valageas00,benson01}). We compute the smoothed
distribution, $y_{\rm s}(\theta)$, for each point in the $(M,z_{\rm
o},z_{\rm f})$ plane and then determine the mean $y_{\rm s}$,
$\bar{y}$, and its variance, $\sigma^2$, for each cluster, averaged
over the whole sky. Assuming the positions of the clusters on the sky
to be uncorrelated (a good approximation as discussed in
\S\ref{sec:cosvar}), the central limit theorem implies that, when the
contributions from all clusters are added together, the observed mean
background will be $y_{\rm bg}=\sum_i \bar{y}_i$ with variance
$\sigma_{\rm bg}=\sum_i \sigma^2_i$, where the sums are taken over all
clusters\footnote{Since the distribution of $y$ is different for each
cluster the central limit theorem only applies under certain
conditions (see, for example, \citealt{eadie}), but we have checked
that these apply for this particular calculation.}.

We take the background to be $y_{\rm bg}+3\sigma_{\rm bg}$ since below
this any signal will be lost in noise due to the unresolved halos. The
computed values of the background will differ between experiments,
since the value of $\sigma_{\rm bg}$ depends upon the beam width, and
also depend upon the cosmological model. (Note that our values of
$y_{\rm bg}$ are consistent with the upper limit of $1.5\times
10^{-5}$ from the {\sc firas} instrument; \citealt{fixsen96}.)
Typically we find backgrounds in the $\tau$CDM cosmology to be
approximately half those in the low-density models (comparable to the
results of \citealt{dasilva00}). Since the effective background will
depend to some degree on the particular analysis method applied to
data we choose to adopt a single value of the background for each
experiment (i.e. the background is assumed independent of cosmology)
to permit easier comparison between models. These values are listed in
Table~\ref{tb:expparams}.

\section{Results}
\label{sec:results}

We present results tuned specifically to the {\sc bolocam} experiment
\citep{glenn98} but will also consider the {\sc planck} experiment
\citep{passvogel00} which will survey a much larger area, albeit with
lower angular resolution than {\sc bolocam}. For our purposes, these
two experiments differ mostly in terms of the solid angle of sky which
they will survey and their beam size which determines the flux of each
cluster, and also influences the background. The specific parameters
used for these two experiments are listed in Table~\ref{tb:expparams}.

We will also consider three different cosmological models,
$\Lambda$CDM (which is currently the most favoured observationally),
$\tau$CDM and OCDM. The default parameters of these models are given
in Table~\ref{tb:cospars} and are used for all calculations unless
otherwise noted. We choose to fix the value of $\Omega_{\rm b} h^2$
(we define the Hubble constant to be $H_0=100 h$km/s/Mpc) to $0.019$
based on the results of \cite{burles98}. The value of $\sigma_8$ is
fixed from the observed abundance of clusters \citep{ecf96}, and the
spectral shape parameter, $\Gamma$, is chosen to match the shape of
galaxy power spectra on large scales \citep{2dfpk}. All the models
have a primordial power spectrum with spectral index $n=1$.

\begin{table}
\caption{Parameters of the cosmological models considered in this work.}
\label{tb:cospars}
\begin{center}
\begin{tabular}{lcccccc}
Model & $\Omega_0$ & $\Lambda_0$ & $\Omega_{\rm b}$ & $h$ & $\sigma_8$ & $\Gamma$ \\
\hline
$\Lambda$CDM & 0.3 & 0.7 & 0.0388 & 0.7 & 0.93 & 0.21 \\
$\tau$CDM & 1.0 & 0.0 & 0.0760 & 0.5 & 0.52 & 0.21 \\
OCDM & 0.3 & 0.0 & 0.0388 & 0.7 & 0.87 & 0.21 
\end{tabular}
\end{center}
\end{table}

We now present results for key observable properties of SZ
clusters. We begin by examining the differences between the {\sc
bolocam} and {\sc planck} experiments (\S\ref{sec:resBP}), proceed to
explore the theoretical uncertainties in the results
(\S\ref{sec:resMF}, \S\ref{sec:resform}, \S\ref{sec:resgas}), examine
briefly the dependence on cosmological parameters
(\S\ref{sec:rescosmo}) and finally consider the effects of preheating
(\S\ref{sec:resent}) and non-Gaussianity (\S\ref{sec:nongauss}),
focusing in particular on what constraints can be put on these
mechanisms and also on cosmological parameters.

\subsection{A Case Study: BOLOCAM vs. PLANCK}
\label{sec:resBP}

We begin by briefly contrasting the two experiments that will be
considered in this work. (We show results for the $\Lambda$CDM
cosmology and the J2000 $P(y)$.) Figure~\ref{fig:BvP} shows three
statistics of the SZ clusters that may be seen by the {\sc bolocam}
and {\sc planck} experiments. The left panel is a plot of the number
of clusters expected versus the flux limit. The middle and right
panels are plots of the redshift and angular size distributions
respectively for clusters with $|S_{143{\rm GHz}}|\geq1$mJy. The
redshift distribution is normalized by the total number of clusters
brighter than this flux. (Unless noted otherwise, the format of all
subsequent figures showing SZ survey statistics will be the same.)

It is immediately obvious that the smaller beam of the {\sc bolocam}
experiment allows it to detect many low-flux clusters which would
disappear into the background for {\sc planck}. As a result, {\sc
bolocam} should find over ten times more cluster per square degree
than {\sc planck}, resulting in approximately 10 SZ cluster detections
per square degree. Of course, the full-sky coverage of {\sc planck}
will mean it detects many more clusters (several thousands) in
total. Because of its greater angular resolution, {\sc bolocam} will
detect clusters with a much broader redshift distribution, which will
help discriminate between models with different cosmological
parameters (see \S\ref{sec:constraints}). Finally, the distribution of
cluster sizes also reflects the size of the experimental beam in each
case, with only a small fraction (approximately 10\% for {\sc bolocam}
and {1\% for {\sc planck}) of detected clusters being resolved (as was
noted for the {\sc planck} experiment by \citealt{kay01}).

Given the predicted slope of the SZ cluster counts we find that the
currently proposed {\sc bolocam} survey (1 square degree surveyed to a
limit of 1mJy) is close to optimal for fixed total observation time in
terms of maximizing the number of detections. A slightly shallower and
broader survey would detect somewhat more SZ clusters, but a deeper
and narrower survey has little advantage since the cluster counts
begin to turn over because the small, faint clusters begin to blend
into the background due to the beam size.

Figure~\ref{fig:contribs} shows which regions of the $M_{\rm
halo}$--$z_{\rm f}$ plane are detectable at different observed
redshifts for a $|S_{143{\rm GHz}}|=1$mJy limited survey. Each line in
the figure corresponds to a single $z_{\rm obs}$, and only clusters
with $(M_{\rm halo},z_{\rm f})$ above and to the right of the line
would be detected. It is immediately apparent that the minimum cluster
mass detectable depends only weakly on the observed redshift. The
dependence arises through the angular diameter distance (since the
total SZ flux scales approximately as $d_{\rm A}^{-2}$ in the absence
of instrumental beam effects). The minimum mass for detection is also
seen to be lower for objects with higher $z_{\rm f}$ since these
objects are denser and so produce larger SZ flux for given mass. (The
lines become horizontal at $z_{\rm f}=z_{\rm obs}$ since no cluster
can have formed after it was observed.)

\begin{figure}
\psfig{file=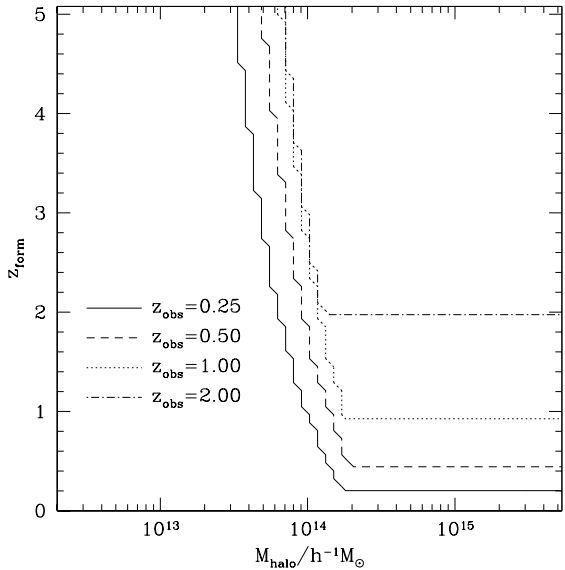,width=80mm}
\caption{The regions of the $M_{\rm halo}$--$z_{\rm f}$ plane in which
clusters can be detected in an $|S_{143{\rm GHz}}|=1$mJy limited {\sc
bolocam} survey. For clusters observed at given $z_{\rm obs}$ (as
indicated by the different line types), only those with $(M_{\rm
halo},z_{\rm f})$ above and to the right of the line are detected.}
\label{fig:contribs}
\end{figure}

\subsection{Assessing the Theoretical Uncertainties}
\label{sec:theorunc}

In this subsection we explore the theoretical uncertainties inherent
in current analytic models of SZ cluster properties. Our aim is to
assess the relative importance of these uncertainties and highlight
where further work is needed to produce an accurate model.

\subsubsection{Sensitivity to the Mass Function}
\label{sec:resMF}

Many previous analytic studies of the SZ effect have used the
Press-Schechter mass function \citep{ps74}, but more accurate fitting
formulaes are now known. To explore the differences intriduced by
these alternative formulae, we have implemented both the
Press-Schechter mass function and two others based upon N-body
simulations developed by J2000 and ST. The J2000 mass function is
currently the best match to numerical simulations over a wide range of
mass scales, and should be taken as giving the best estimate of the
halo mass function. We show results for the other mass functions
simply for comparison.

The results for all three mass functions for the {\sc bolocam}
experiment are displayed in Fig.~\ref{fig:Pydep}. The main differences
between the models show up in the redshift distribution of
clusters. The ST mass function produces the largest fraction of high
redshift clusters while Press-Schechter produces the least (a factor
of 4 fewer than ST for $z > 1$). Changing the mass function also
produces slight variations in the total number of clusters seen. As we
will show (\S\ref{sec:constraints}), an accurate redshift distribution
will be crucial in determining cosmological parameters and details of
the gas physics, in particular for the {\sc bolocam} experiment. For
that experiment, number counts alone do not contain sufficient
information to be able to provide interesting constraints.

\begin{figure*}
\begin{tabular}{c@{}c@{}c}
\psfig{file=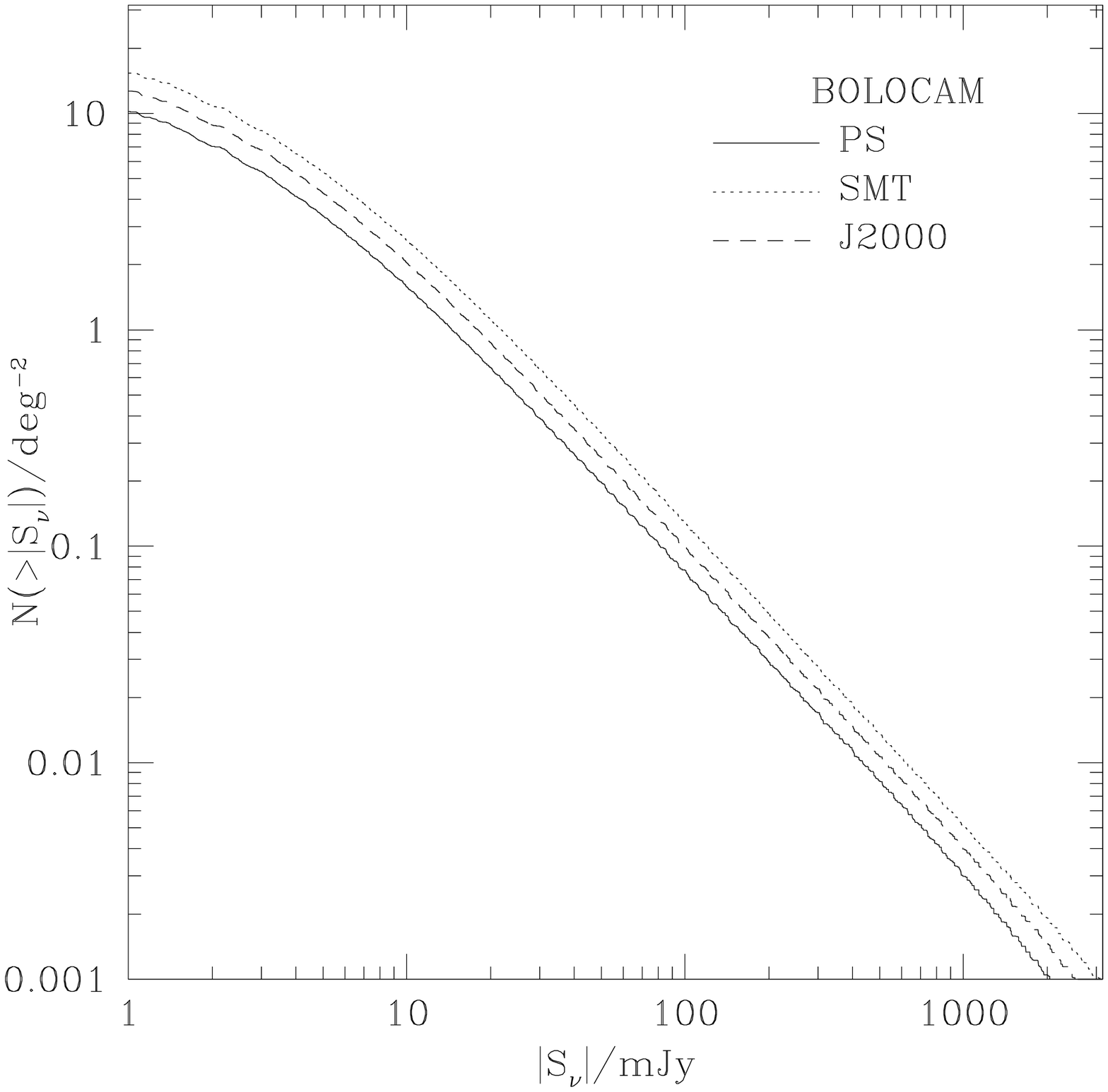,width=58mm} &
\psfig{file=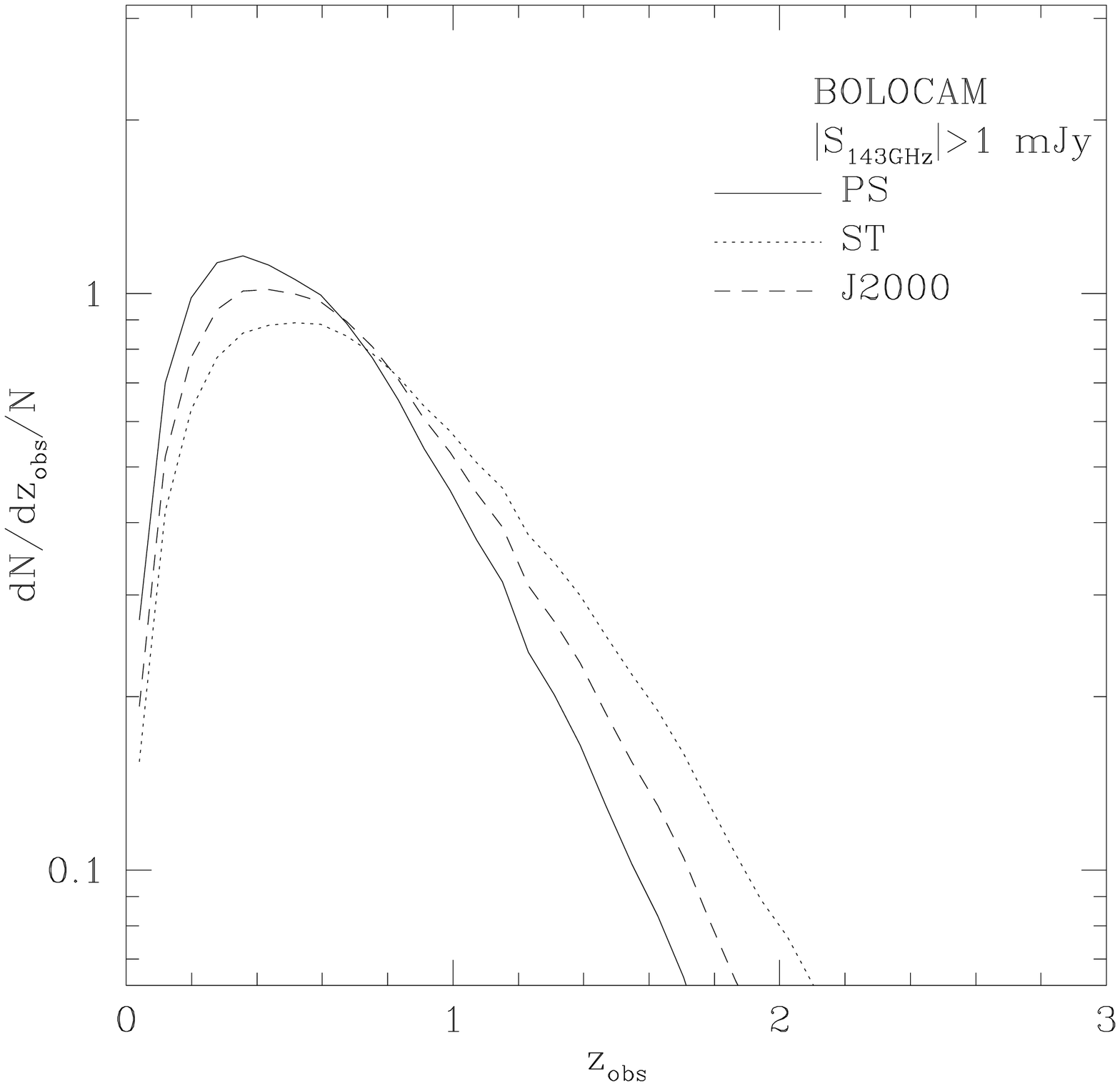,width=58mm} &
\psfig{file=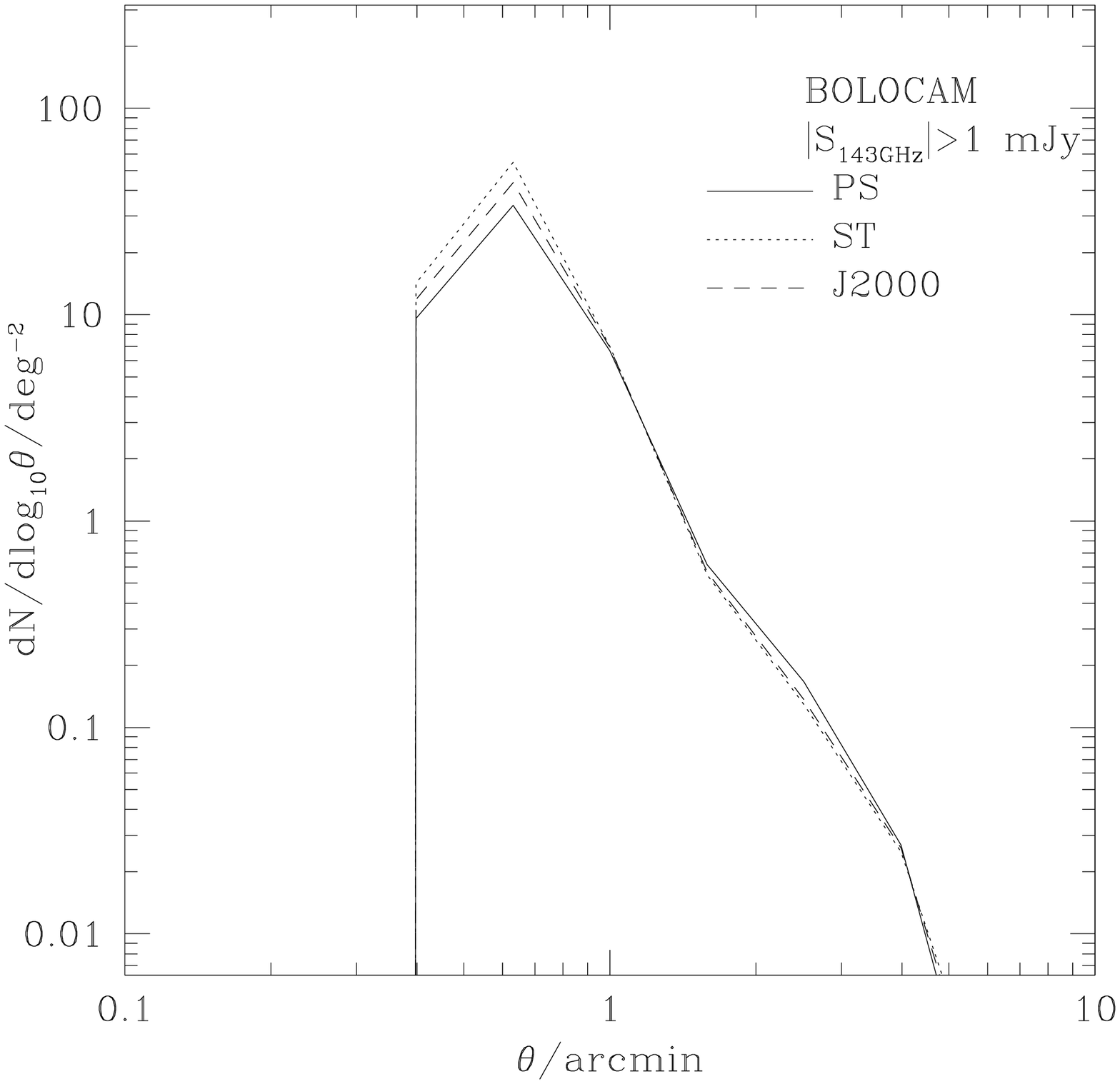,width=58mm}
\end{tabular}
\caption{Properties of SZ clusters for different determinations of
the halo mass function, as characterised by $P(y)$. Solid, dotted and
dashed lines show results for the PS, ST and J2000 forms for $P(y)$
respectively. The left-hand panel shows counts as a function of flux
at 143GHz, the centre panel shows the redshift distribution of
clusters brighter than $|S_{143{\rm GHz}}|=1$mJy, and the right hand
panel shows the distribution of angular sizes for the same
clusters. (All subsequent figures will follow this general format.)
All results are for the $\Lambda$CDM cosmology and the parameters of
the {\sc bolocam} experiment.}
\label{fig:Pydep}
\end{figure*}

The differences between models with these three forms for $P(y)$ as
shown in Fig.~\ref{fig:Pydep} should be compared to the expected
statistical errors from individual experiments. In the case of {\sc
bolocam} the statistical errors (which should be close to
Poissonian---see \S\ref{sec:rescosmo}) will be significantly larger
than the uncertainties due to the form of $P(y)$. For the {\sc
planck} experiment however, the statistical errors will be much
smaller, making the systematic differences due to the form of $P(y)$
the dominant uncertainty.

\subsubsection{Dependence on Formation Redshift Prescription}
\label{sec:resform}

We next examine the effect of varying the prescription for cluster
formation redshifts. The formation redshift determines the cluster's
gas density and pressure, which directly impact the detectability of
the cluster. This is a potentially significant effect on the number
counts of SZ clusters.

At present it is not fully clear how a cluster's formation and merger
histories influence the thermodynamic properties of the gas it
contains. While numerical simulations may soon clarify this issue we
have for now explored three possibilities for the formation redshifts
of clusters (which we assume fix the thermodynamic properties of the
gas as described in \S\ref{sec:model}).

Comparing the predictions of the three formation redshift formulas in
Fig.~\ref{fig:form}, we find that the functions proposed by
\cite{lc93} and \cite{sasaki94} yield quite similar results. However,
the lower limit of $z_{\rm f} = z_{\rm o}$ yields substantially
smaller number counts. Interestingly, all three models produce
remarkably similar normalised redshift distributions. It is worth
noting that the differences in number counts for our three
formation-redshift distributions are greater than the differences
resulting from the various forms for $P(y)$ discussed in
\S\ref{sec:resMF}, and at 1mJy become comparable to the random errors
expected in the {\sc bolocam} experiment. These differences highlight
the need for a better understanding of how a cluster's gas properties
are determined in order to make sufficiently accurate calculations of
the abundance of SZ clusters, and to allow cosmological and
gas-distribution parameters to be determined without systematic
biases.

\begin{figure*}
\begin{tabular}{c@{}c@{}c}
\psfig{file=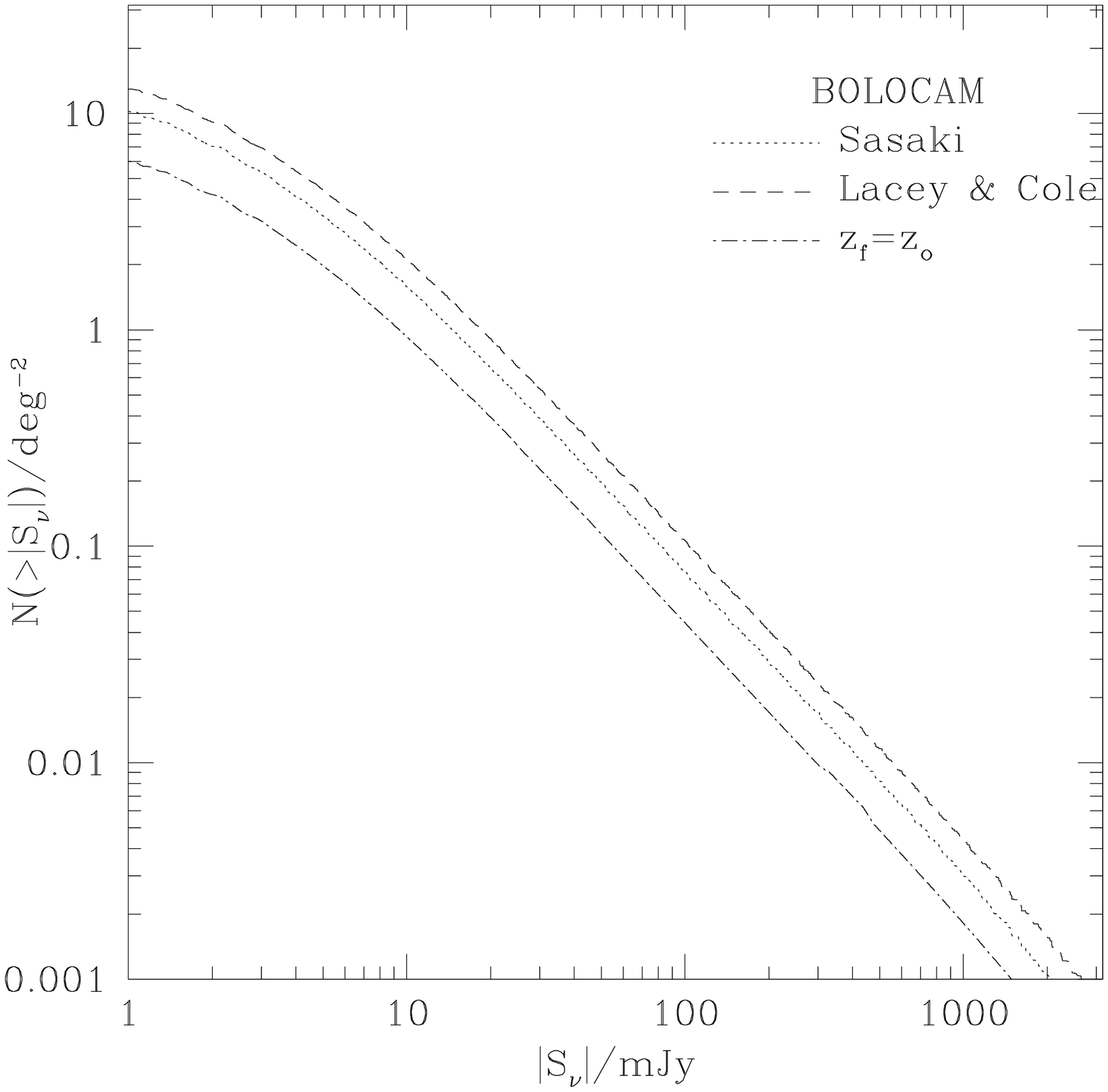,width=58mm} &
\psfig{file=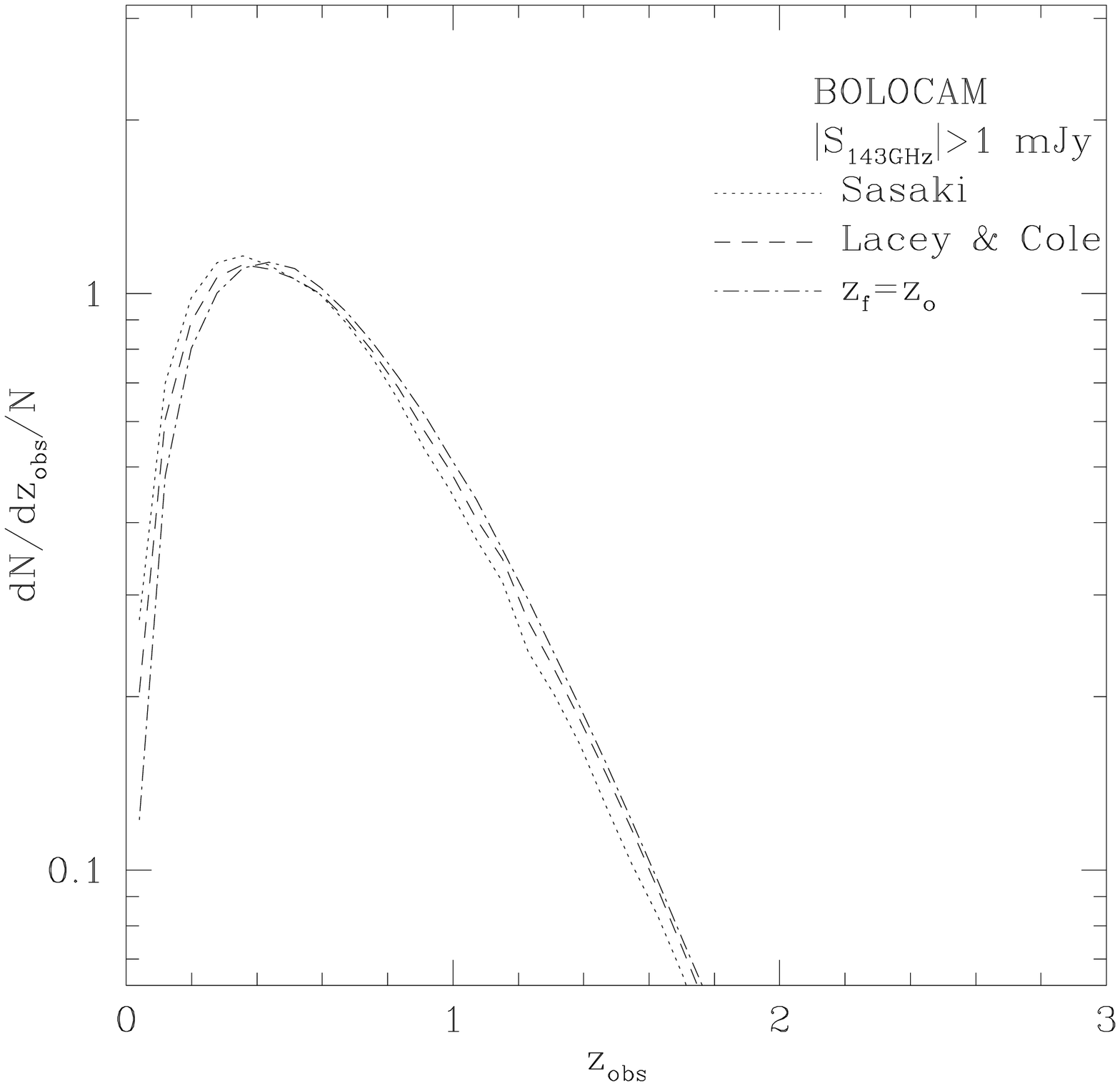,width=58mm} &
\psfig{file=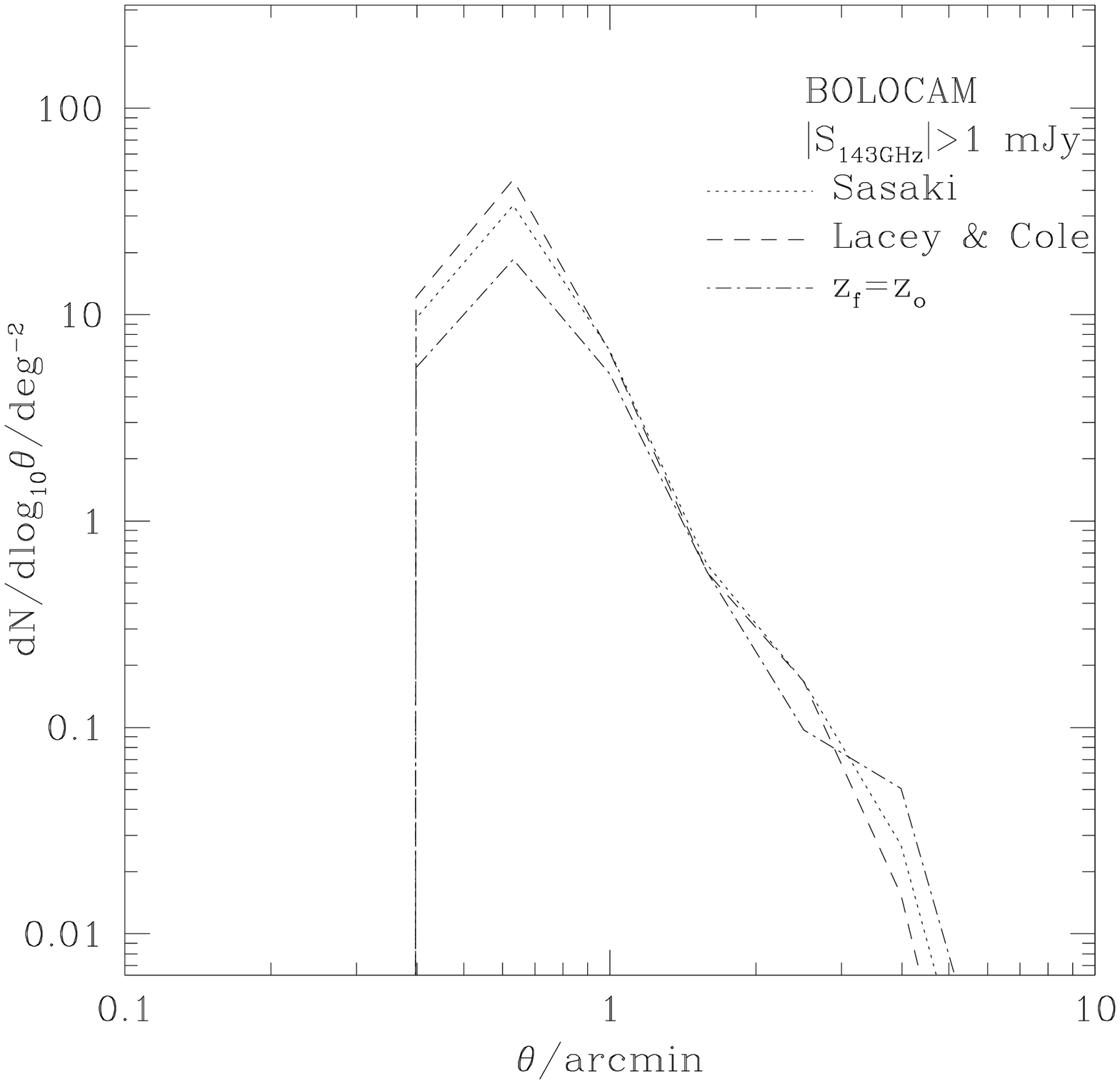,width=58mm}
\end{tabular}
\caption{Properties of SZ clusters for different formation redshift
calculations. Dotted and dashed lines show results for the
\protect\citet{sasaki94} and \protect\citet{lc93} formation redshift
distributions respectively, while the dot-dashed line indicates the
results when halos are assumed to have formed at the redshift they are
observed at. The left-hand panel shows counts as a function of flux at
143GHz, the centre panel shows the redshift distribution of clusters
brighter than $|S_{143{\rm GHz}}|=1$mJy, and the right hand panel shows
the distribution of angular sizes for the same clusters. All results
are for the $\Lambda$CDM cosmology and the parameters of the {\sc
bolocam} experiment and use the J2000 $P(y)$.}
\label{fig:form}
\end{figure*}

\subsubsection{Dependence on Gas Temperature Profile}
\label{sec:resgas}

\begin{figure*}
\begin{tabular}{c@{}c@{}c}
\psfig{file=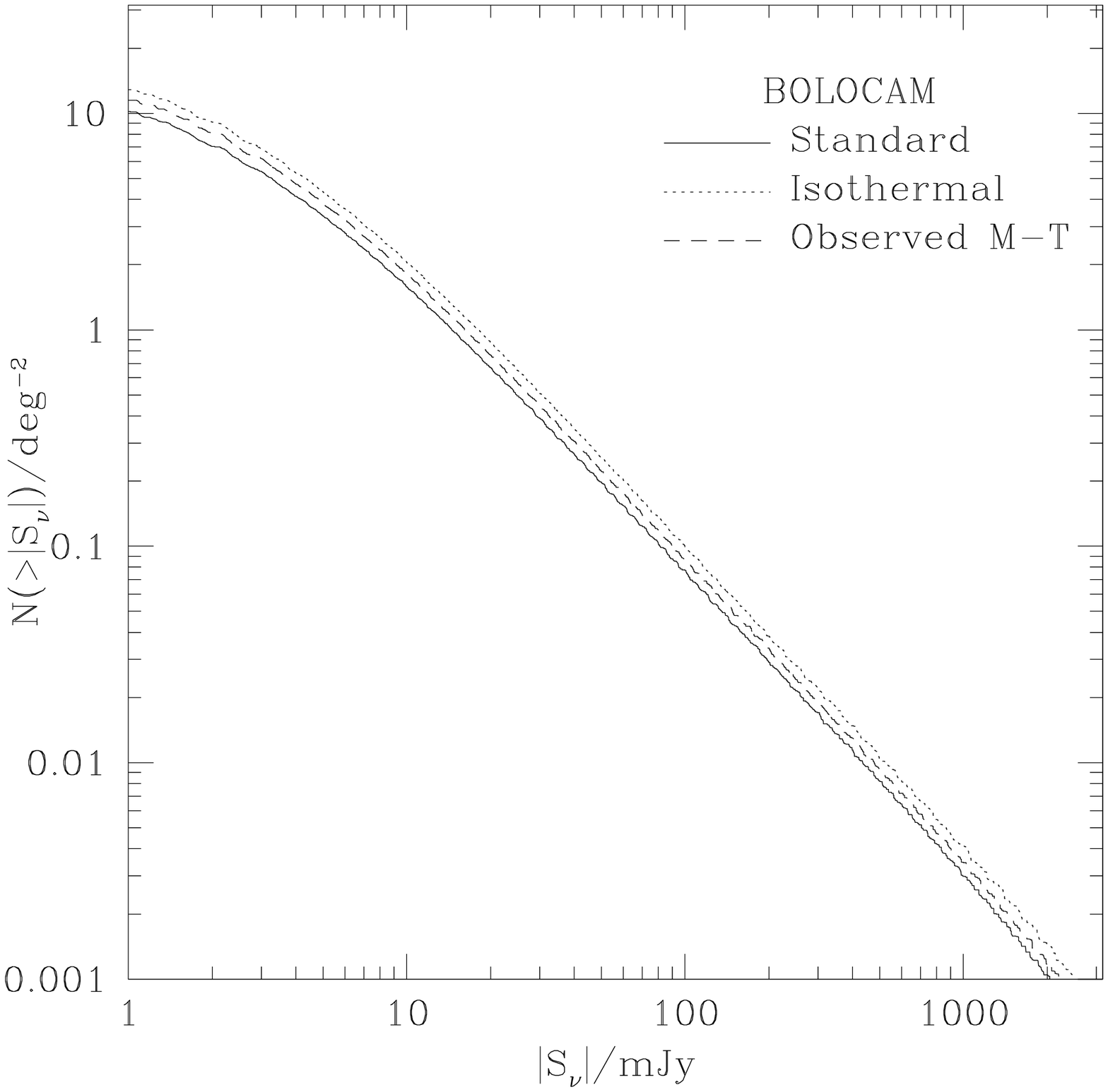,width=58mm} &
\psfig{file=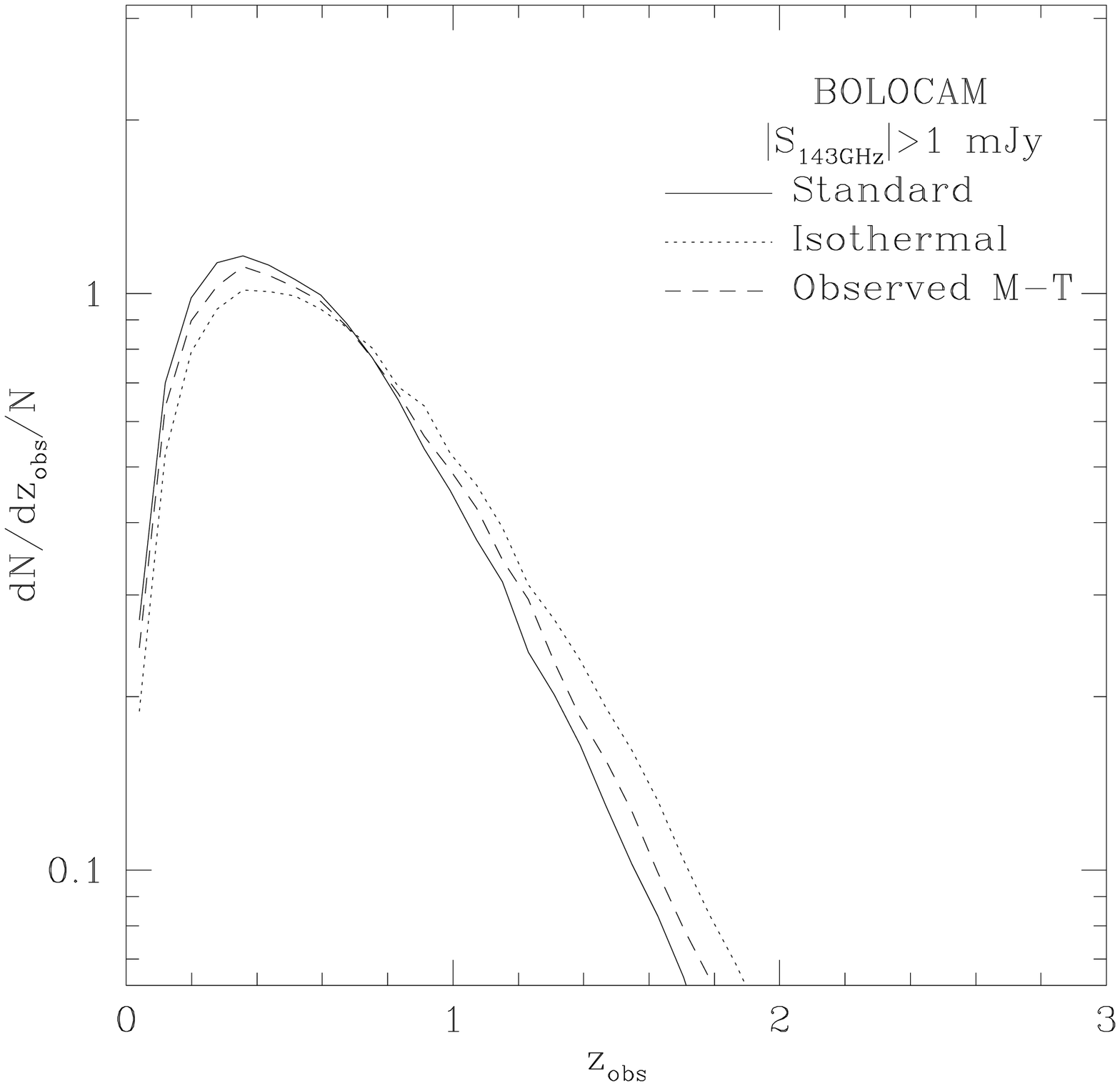,width=58mm} &
\psfig{file=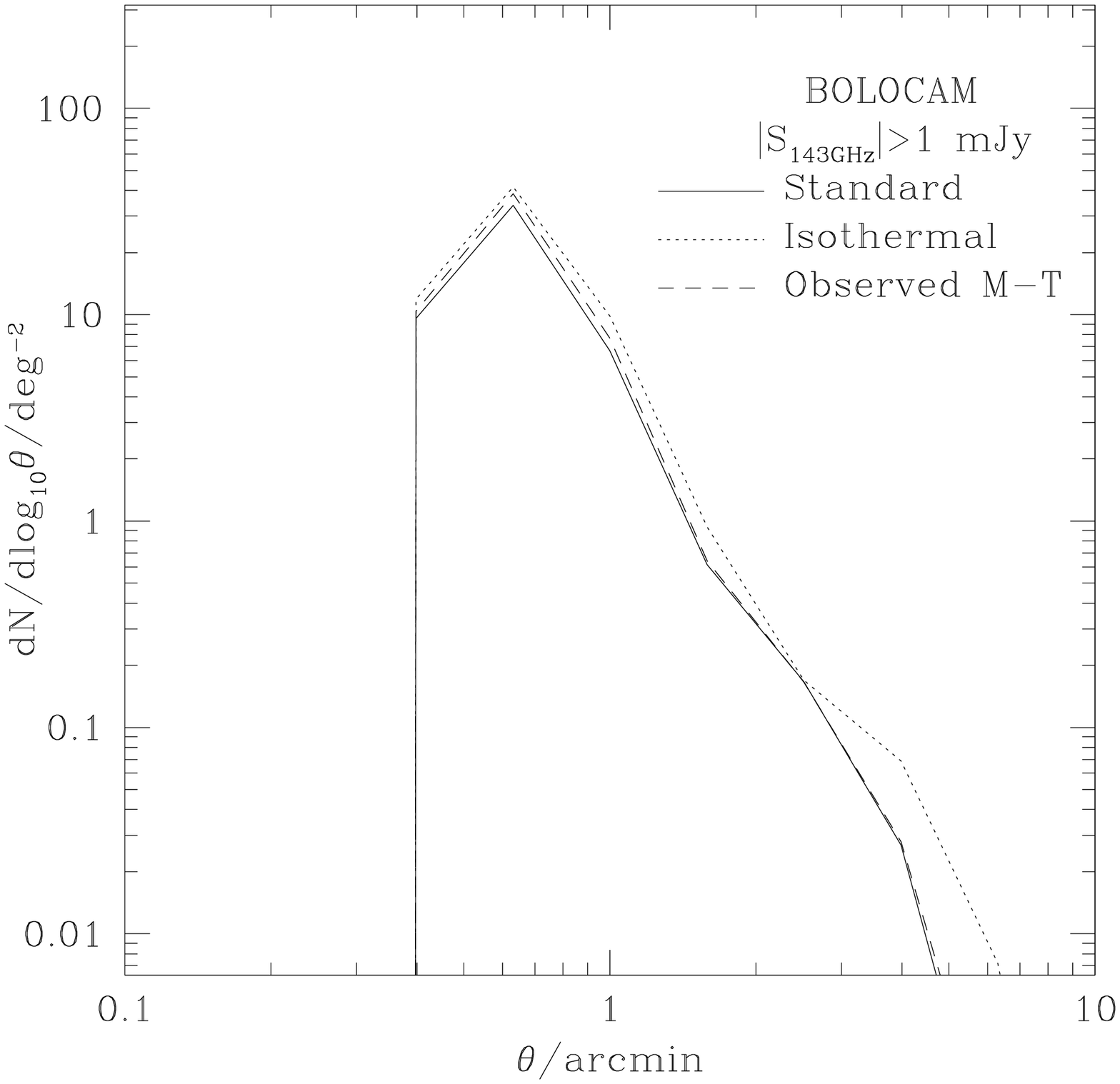,width=58mm}
\end{tabular}
\caption{Properties of SZ clusters for different assumptions about
the temperature profile of cluster gas. Solid, dashed and dotted lines
indicate our standard assumptions, isothermal halos and halos given a
temperature from the observed $M$--$T$ relation respectively. The
left-hand panel shows counts as a function of flux at 143GHz, the
centre panel shows the redshift distribution of clusters brighter than
$|S_{143{\rm GHz}}|=1$mJy, and the right hand panel shows the
distribution of angular sizes for the same clusters. All results are
for the $\Lambda$CDM cosmology and the parameters of the {\sc bolocam}
experiment and use the J2000 $P(y)$.}
\label{fig:gas}
\end{figure*}

The form of the assumed gas density and temperature profiles will
affect the calculation of the SZ flux --- here we examine three
temperature profiles. (Note that \citet{kay01} explored the effect of
changing the gas density profile, finding that a beta-model profile as
used in this work produces SZ fluxes around 40\% higher than an NFW
gas profile.) In our standard case, the outer boundary temperature was
set by the virial theorem, the density set to a beta-model, and the
cluster's temperature profile calculated assuming hydrostatic
equilibrium (as described in \S\ref{sec:nopre}). Since the resulting
mass-temperature relation disagrees with observations
\citep{horner99}, we also considered a model in which the observed
mass-temperature relation was used to determine the outer boundary
temperature with the temperature profile again calculated from
hydrostatic equilibrium. Specifically, we used the relation given by
\citet{kay01} which was based upon the observational data of
\citet{horner99}. Finally, the sensitivity of the SZ flux to the
temperature profile was tested by using an isothermal profile. While
this does not result in hydrostatic equilibrium, it should provide a
useful check as to the sensitivity of the SZ flux to the temperature
profile. In this case the temperature of the gas was set equal to the
virial temperature of the cluster.

As seen in Fig.~\ref{fig:gas}, the effects of these different
temperature profiles are very small, about 20\% (i.e. significantly
smaller than the uncertainties due to $P(y)$ or the formation redshift
distribution, and less than the 40\% differences found by \cite{kay01}
for different gas density profiles), a consequence of the fact that
the temperature varies only slightly with radius even in the
hydrostatic equilibrium profiles. Nevertheless, improvement in this
area would be necessary to fully exploit the high statistical accuracy
of SZ cluster counts obtainable by the {\sc planck} experiment.

\subsection{Constraints on Cosmological Parameters, Non-Gaussianity and Preheating}
\label{sec:constraints}

\subsubsection{Dependence on Cosmological Parameters}
\label{sec:rescosmo}

It is well known that the abundance and redshift distribution of SZ
clusters is sensitive to cosmological parameters as was discussed in
our Introduction. Here we will briefly examine results for different
cosmological parameters to demonstrate how SZ survey statistics are
affected by these quantities, and then explore the constraints that
may be obtained in \S\ref{sec:nongauss} and \S\ref{sec:resent}.

We consider three different sets of cosmological parameters (defined
in Table~\ref{tb:cospars}), $\tau$CDM, $\Lambda$CDM, and, for
completeness, OCDM which is now strongly ruled out by CMB measurements
(e.g. \citealt{debernardis01}). Fig.~\ref{fig:cosmo} shows that there
are large differences in both the total number counts of clusters and
in the redshift distribution between $\tau$CDM and $\Lambda$CDM. As
noted before, the large differences in the redshift distributions
imply that obtaining redshifts for SZ detected clusters will be very
valuable in helping to constrain cosmological parameters. The
errorbars in the left-hand panel of Fig.~\ref{fig:cosmo} show the rms
scatter obtained from our cosmic variance calculations. As expected,
the scatter is very close to Poissonian since the wide SZ redshift
distribution projects out most of the clustering signal.

The most significant difference is in the redshift distribution. Even
a few clusters at $z > 1$ will rule out $\Omega_0 = 1$ in the absence
of significant non-Gaussianity, providing some prior constraint on
$\sigma_8$ is adopted (since even with redshifts the constraints on
$\Omega_0$ and $\sigma_8$ are highly degenerate, see
\S\ref{sec:nongauss}). The results are also sensitive to
$\sigma_8$. Varying $\sigma_8$ slightly scales the number counts up
and down significantly, and alters the high-$z$ tail of the redshift
distribution.
	
\begin{figure*}
\begin{tabular}{c@{}c@{}c}
\psfig{file=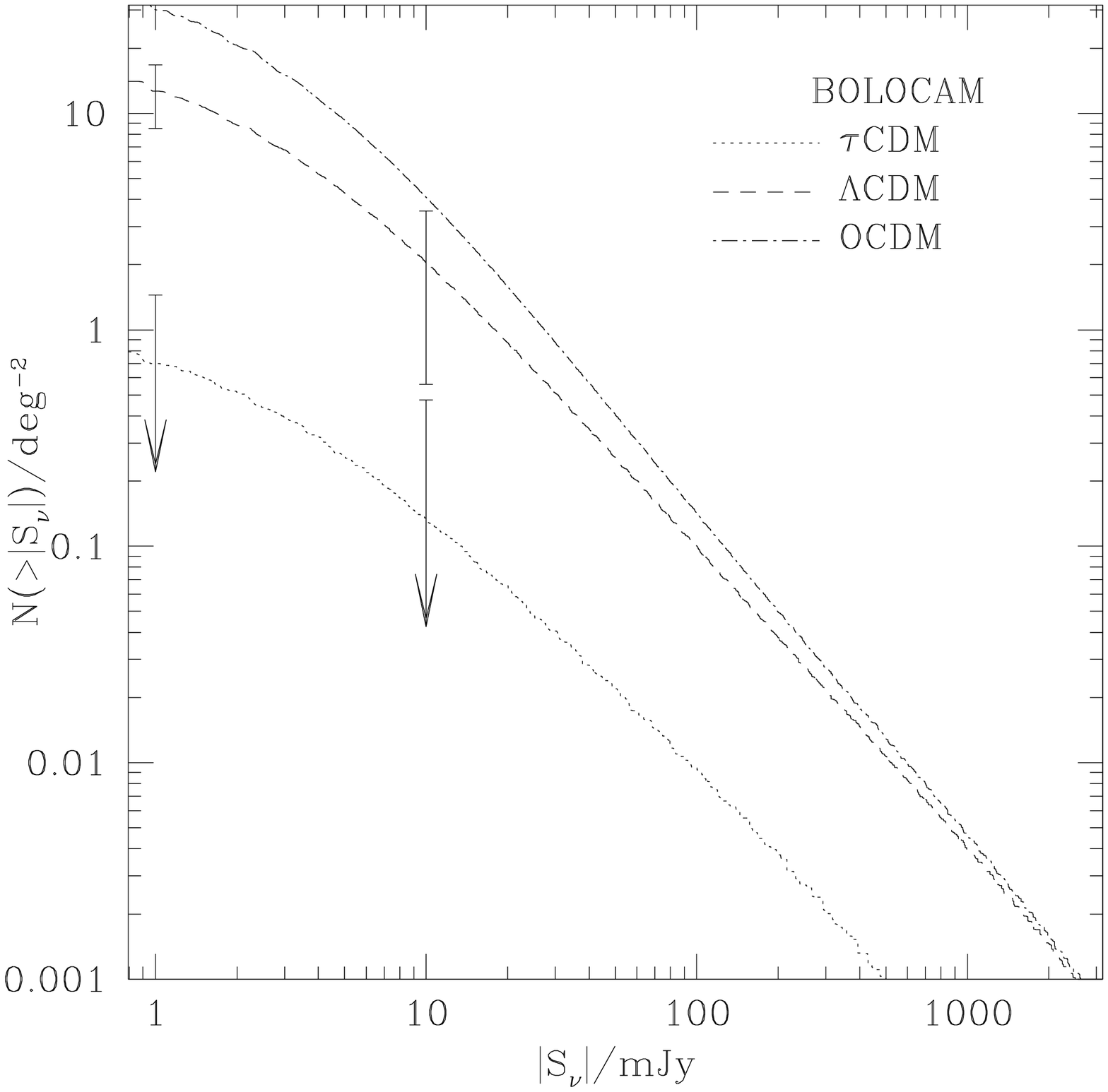,width=58mm} &
\psfig{file=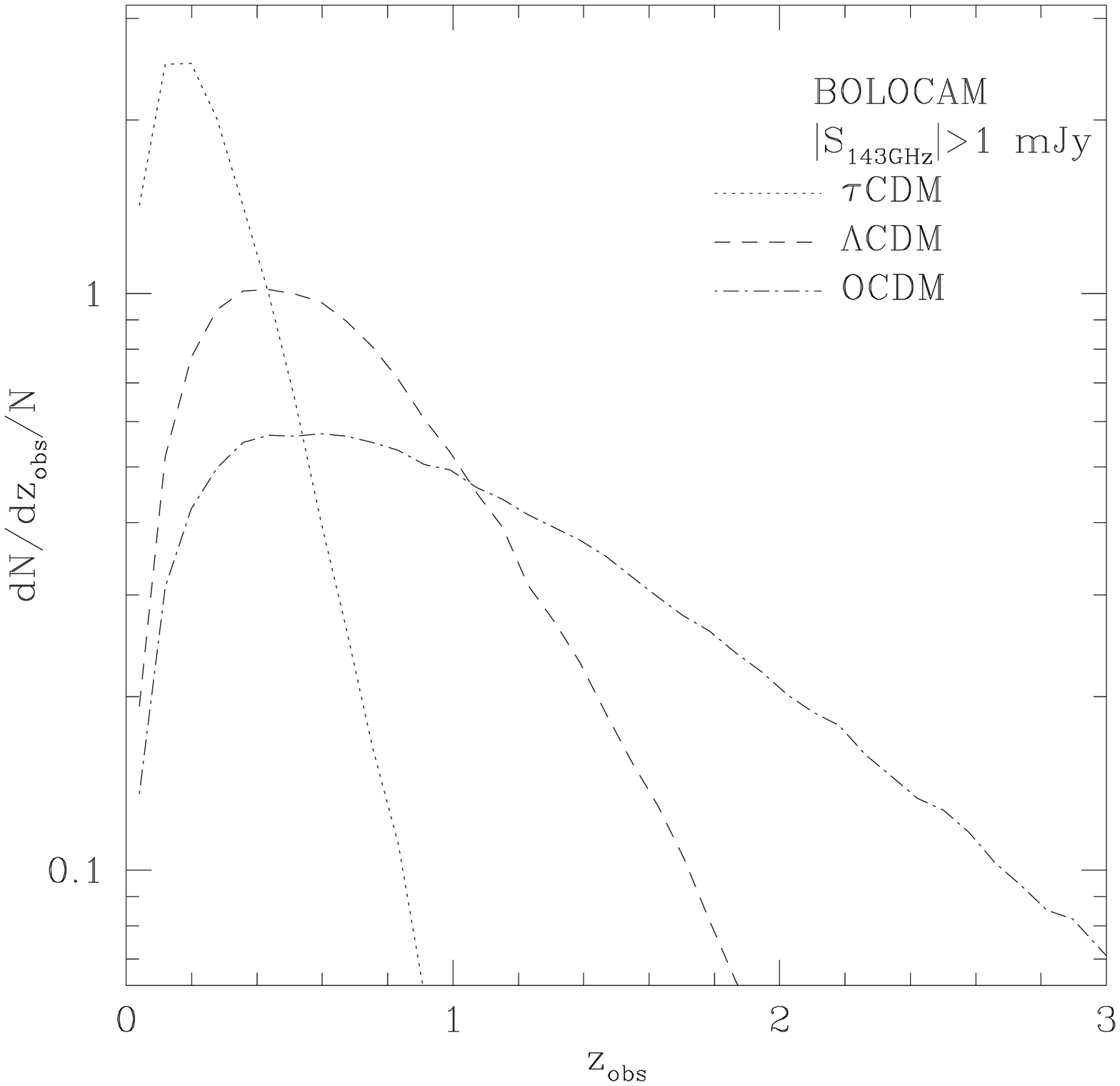,width=58mm} &
\psfig{file=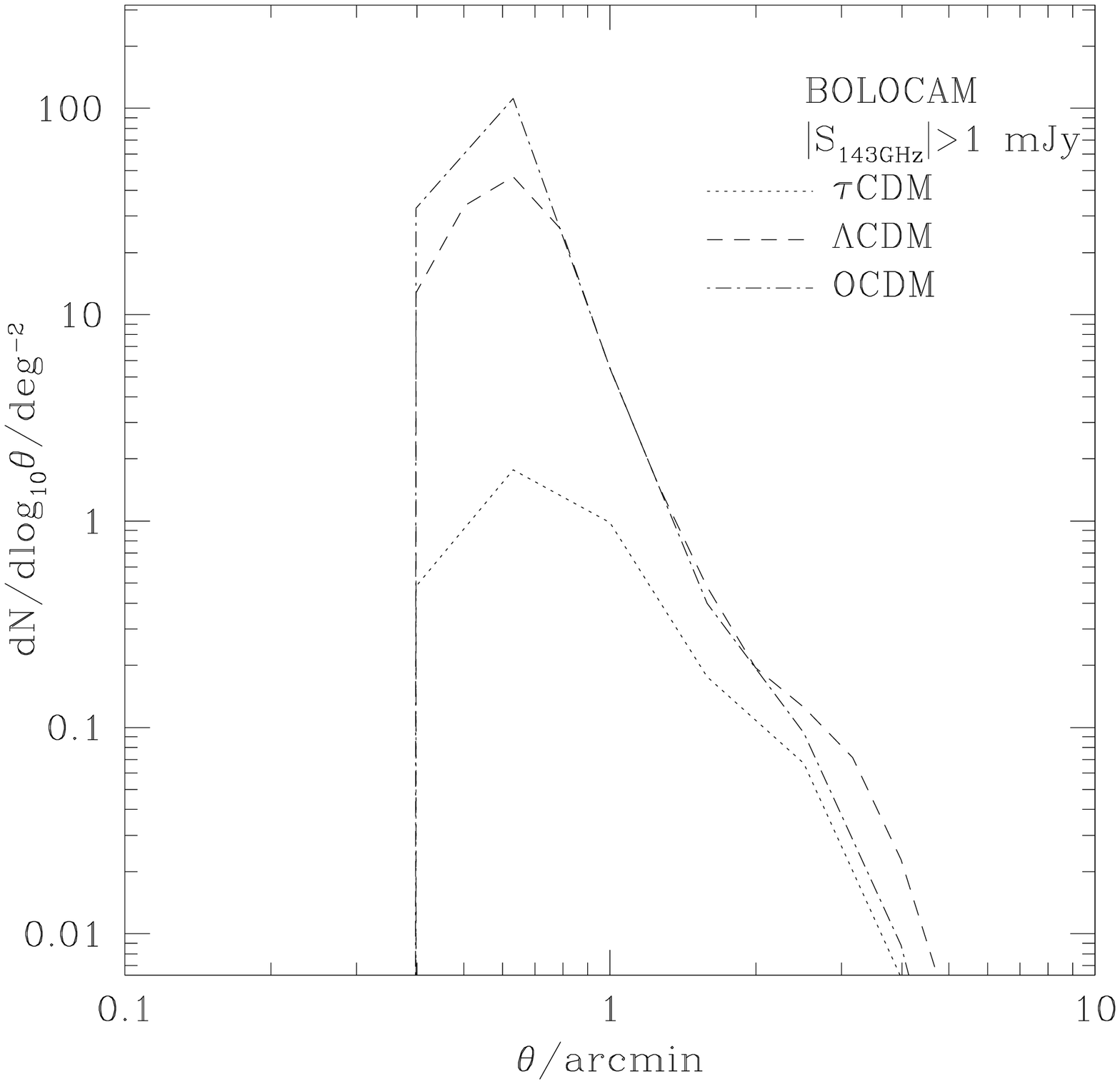,width=58mm}
\end{tabular}
\caption{Properties of SZ clusters for different sets of cosmological
parameters. Dotted and dashed lines show results for $\tau$CDM and
$\Lambda$CDM respectively. The left-hand panel shows counts as a
function of flux at 143GHz (error bars indicate the expected cosmic
variance in 1 square degree fields for the $\tau$CDM and $\Lambda$CDM
cosmologies), the centre panel shows the redshift distribution of
clusters brighter than $|S_{143{\rm GHz}}|=1$mJy, and the right hand
panel shows the distribution of angular sizes for the same
clusters. All results are for the parameters of the {\sc bolocam}
experiment and the J2000 $P(y)$.}
\label{fig:cosmo}
\end{figure*}

\begin{figure*}
\begin{tabular}{c@{}c@{}c}
\psfig{file=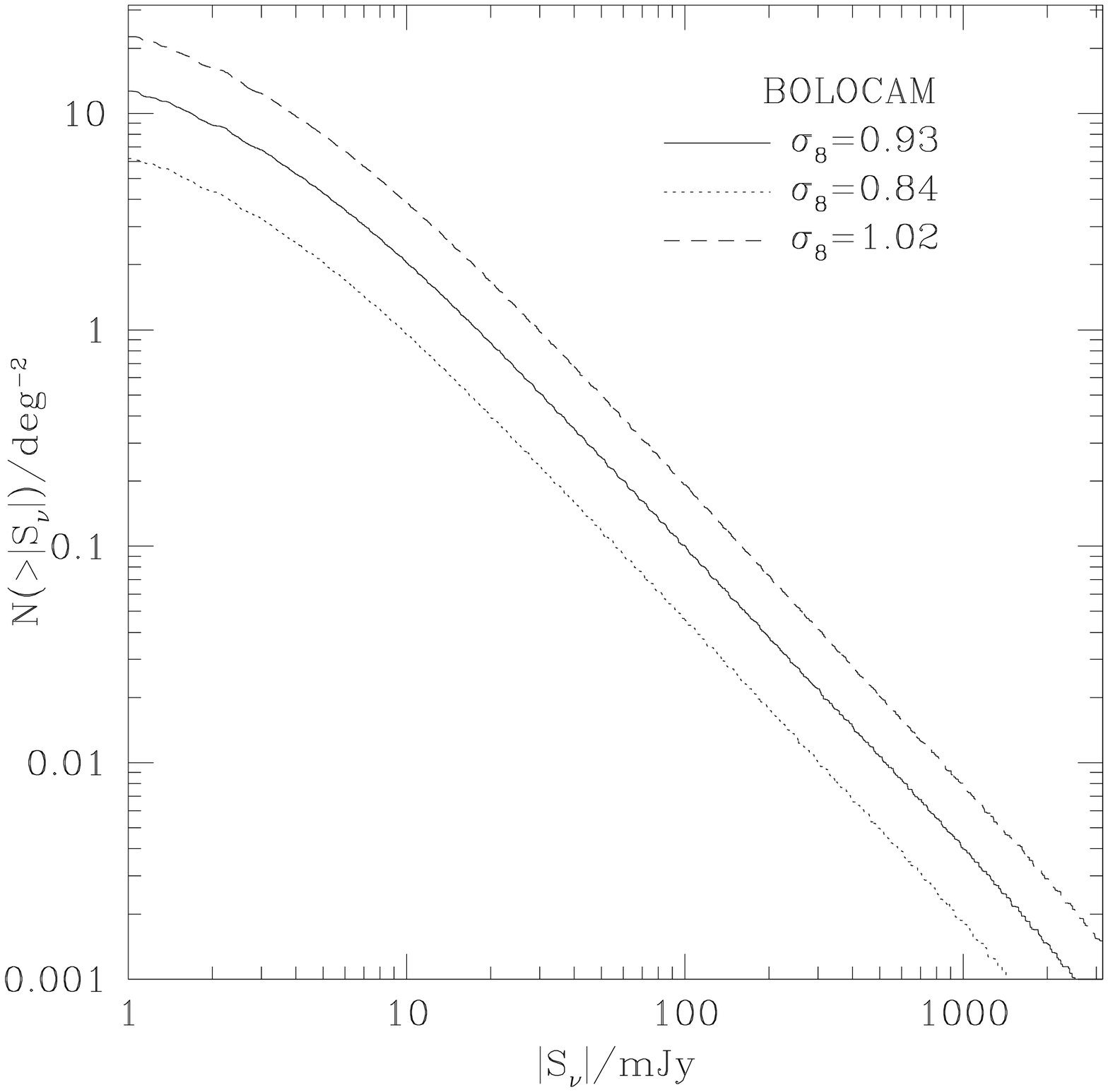,width=58mm} &
\psfig{file=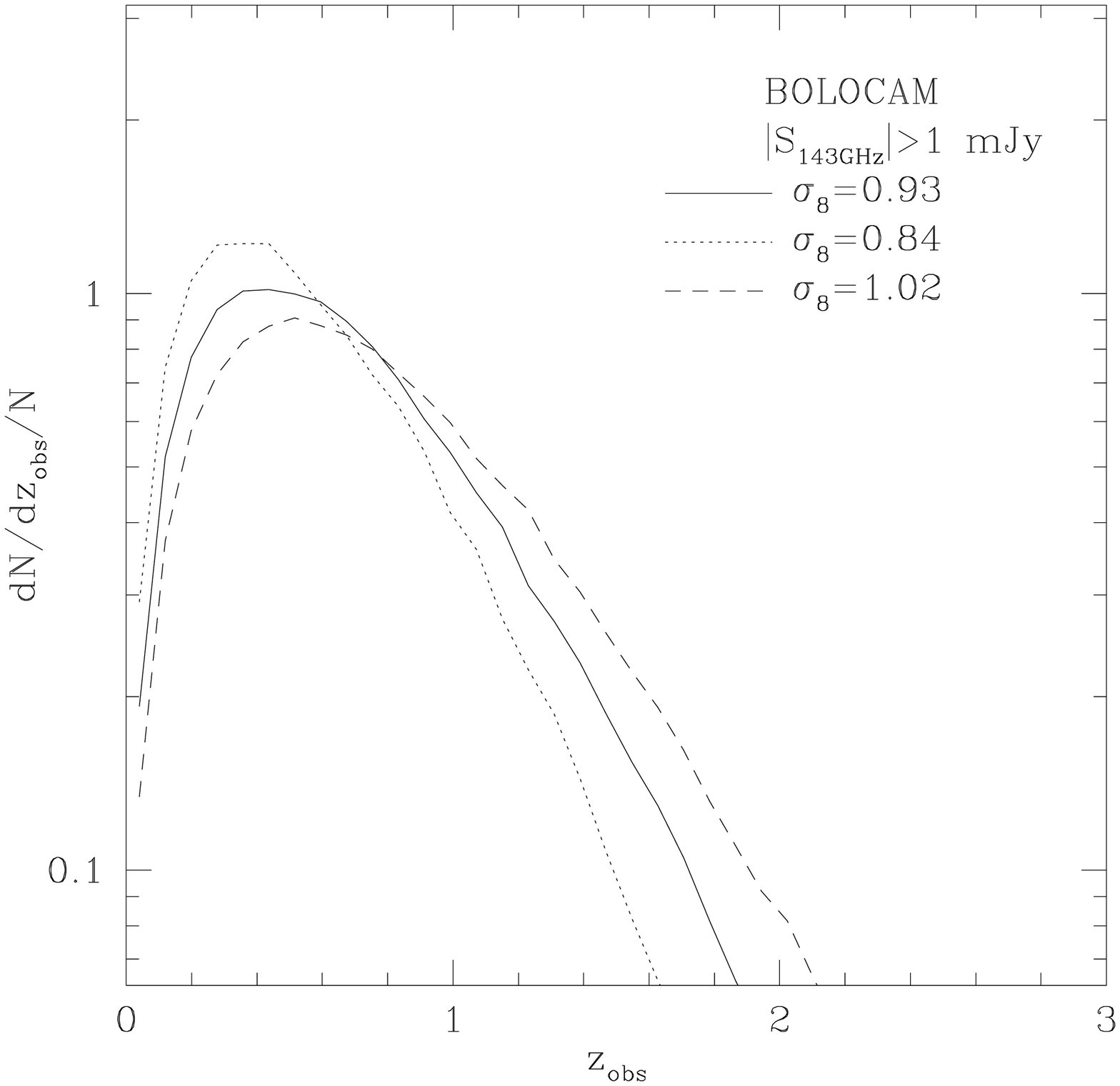,width=58mm} &
\psfig{file=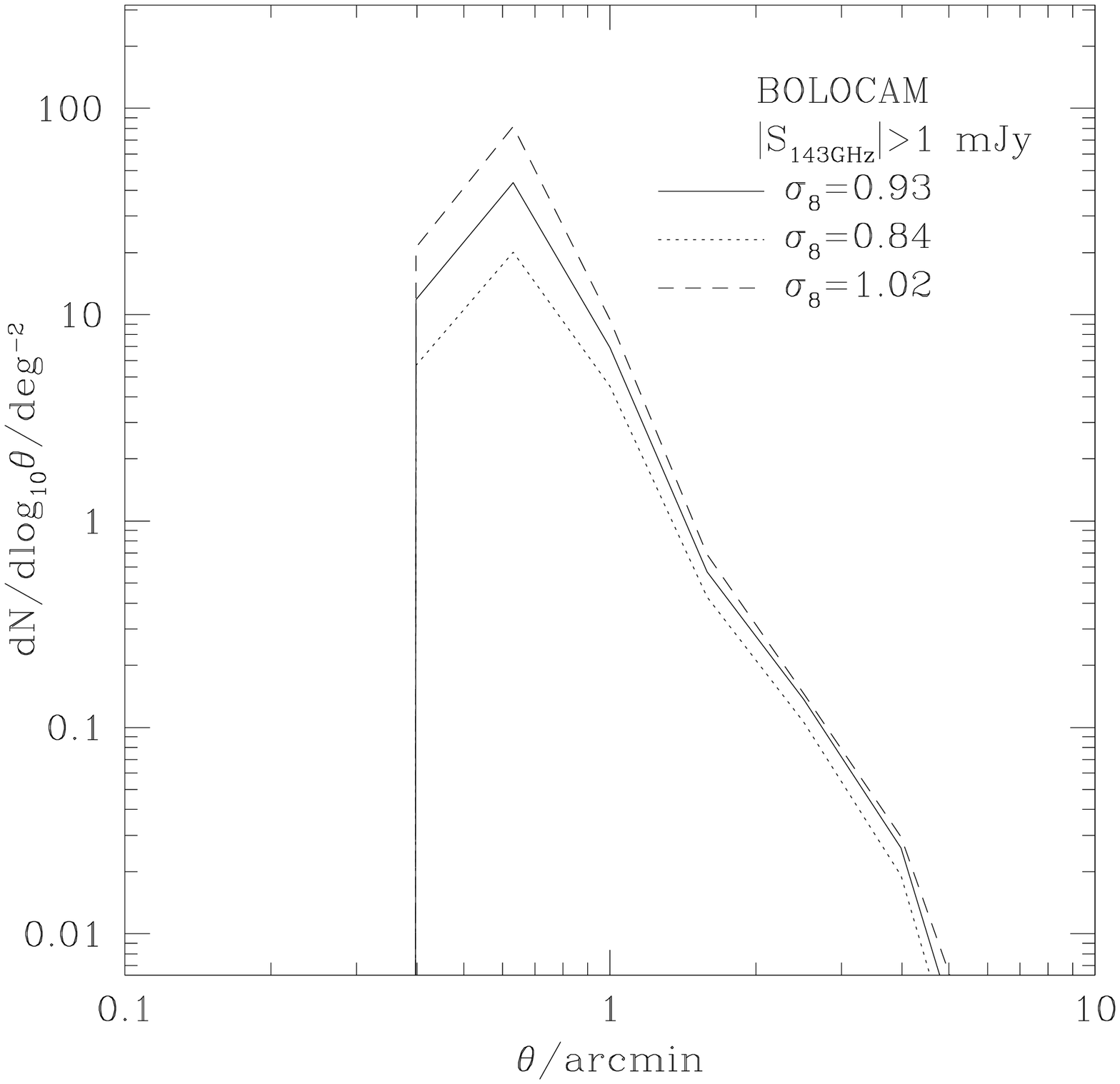,width=58mm}
\end{tabular}
\caption{Properties of SZ clusters for different values of $\sigma_8$
as indicated in the figure. The left-hand panel shows counts as a
function of flux at 143GHz, the centre panel shows the redshift
distribution of clusters brighter than $|S_{143{\rm GHz}}|=1$mJy, and
the right hand panel shows the distribution of angular sizes for the
same clusters. All results are for the parameters of the {\sc bolocam}
experiment and the J2000 $P(y)$.}
\label{fig:sigma8}
\end{figure*}

The observed SZ flux will also depend on the value of $\Omega_{\rm
b}$. In our model, where the effects of radiative cooling are ignored,
the electron density (and hence $y(\theta)$ and $y_{\rm s}(\theta)$)
in each dark matter halo scales in proportion to $\Omega_{\rm
b}$. Furthermore, the background which our model predicts also scales
in proportion to $\Omega_{\rm b}$. As a result of these two scalings
$\theta_{\rm b}$ in eqn.~(\ref{eq:intflux}) is independent of
$\Omega_{\rm b}$ and so the observed, integrated SZ flux of a model
cluster simply scales in proportion to $\Omega_{\rm b}$.

\subsubsection{Constraints on Non-Gaussianity}
\label{sec:nongauss}

As shown in Fig.~\ref{fig:nongauss}, non-Gaussian initial conditions
will have a large effect on both $\d N/\d z$ and $N(>S)$. Cluster
sizes are mostly unaffected. These effects are degenerate with changes
due to varying several of the cosmological parameters. We will now
explore quantitatively these degeneracies, and the constraints that
can be placed on $G$ and various cosmological parameters.

\begin{figure*}
\begin{tabular}{c@{}c@{}c}
\psfig{file=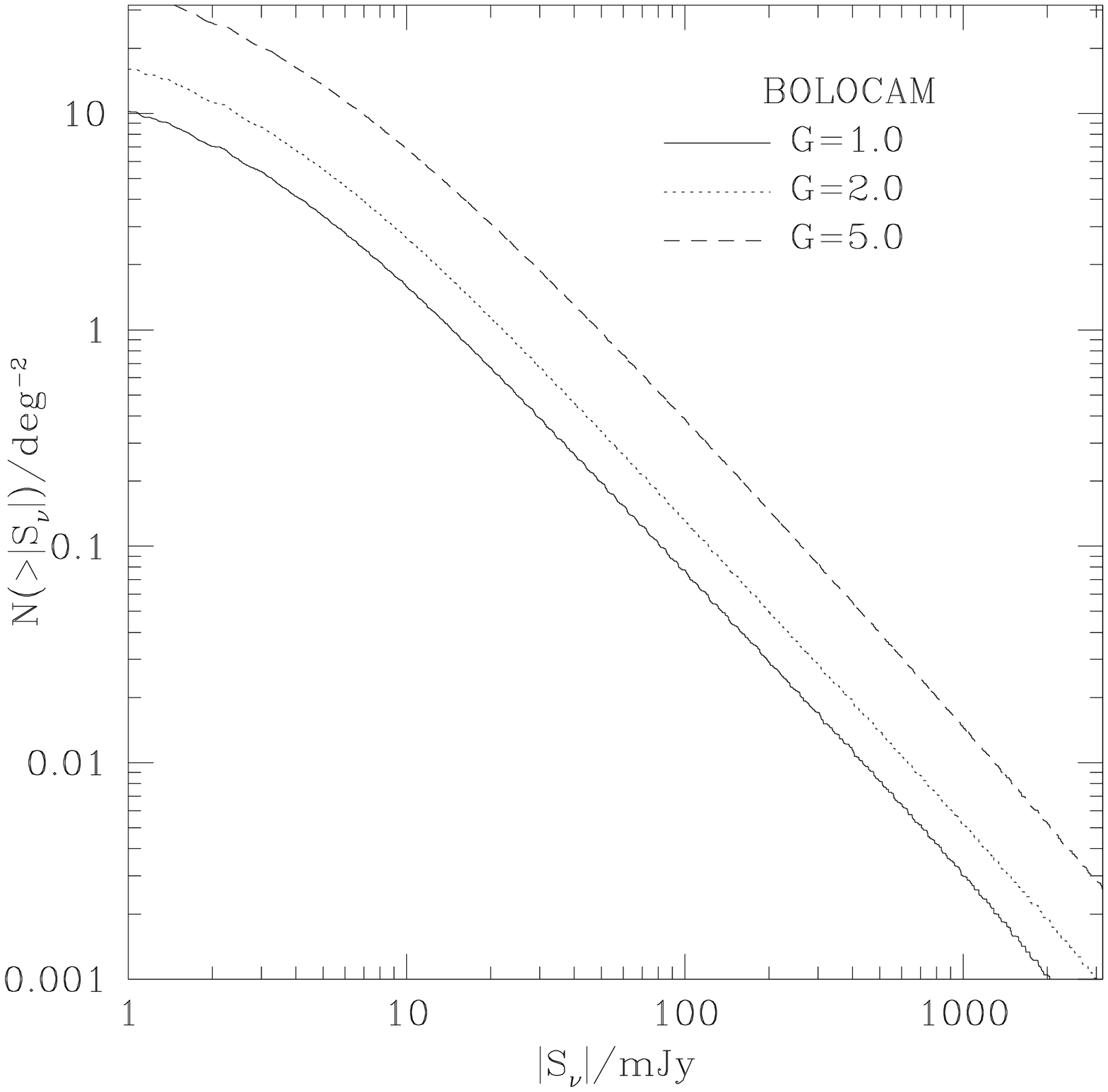,width=58mm} &
\psfig{file=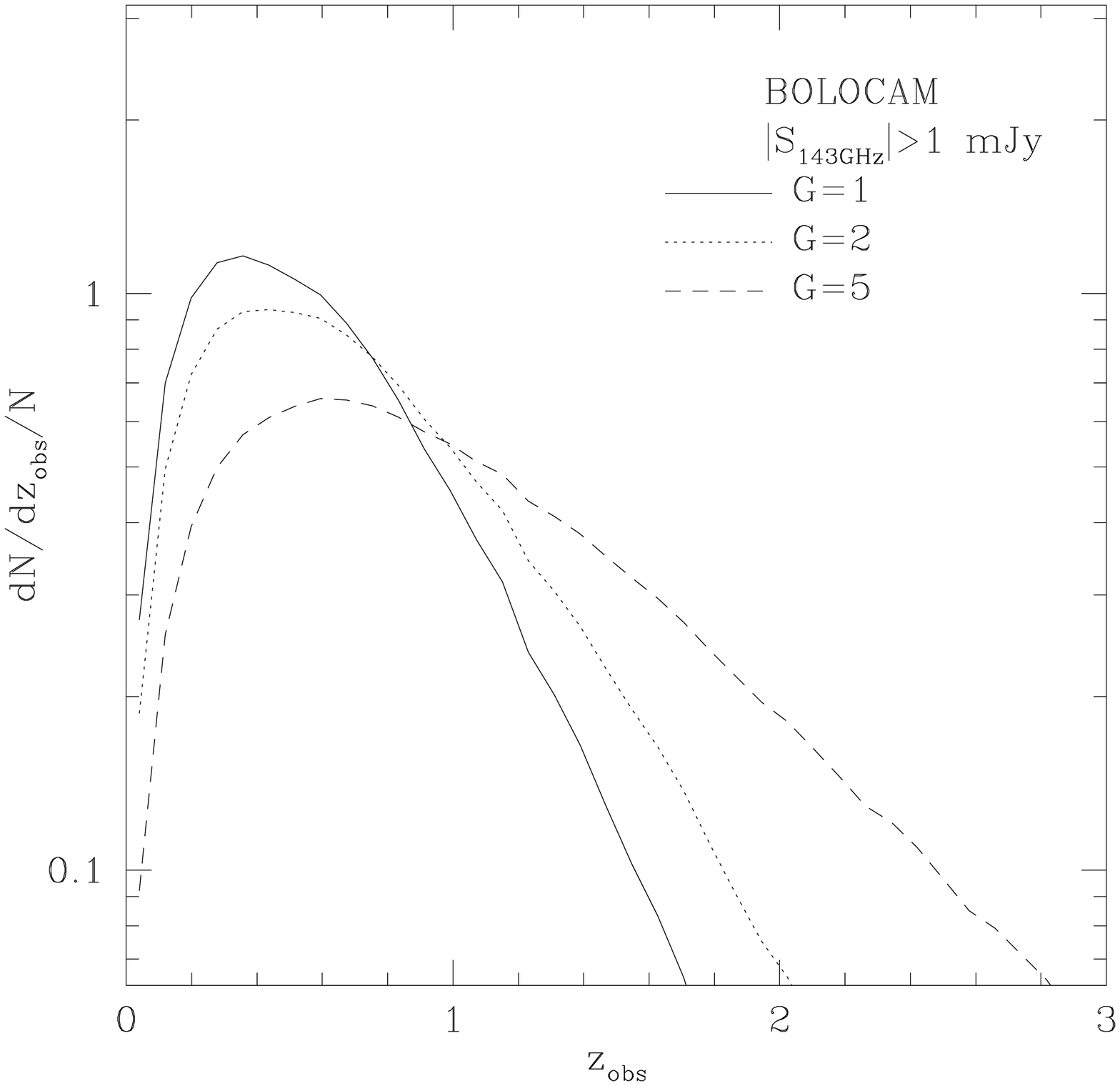,width=58mm} &
\psfig{file=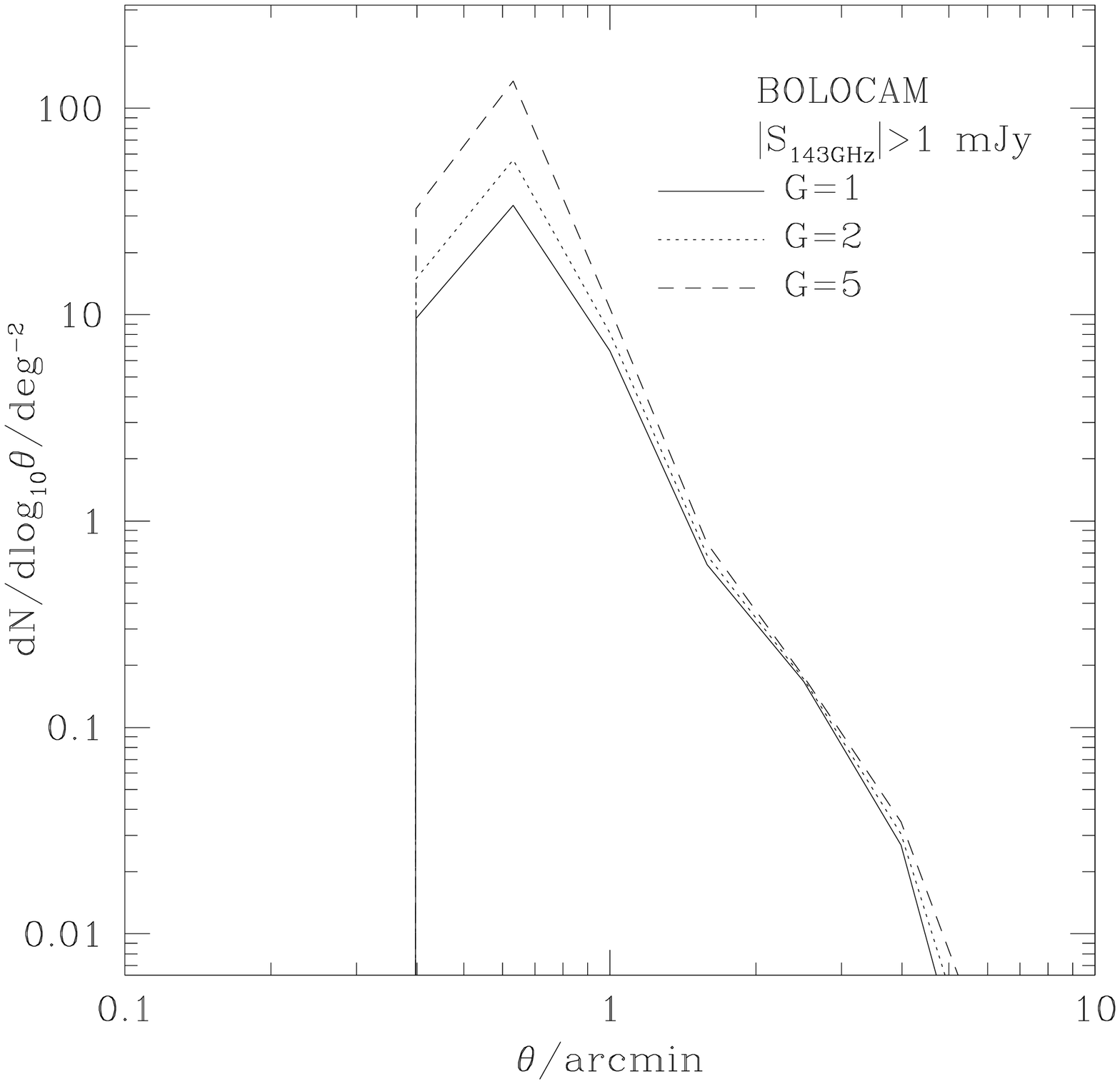,width=58mm}
\end{tabular}
\caption{Properties of SZ clusters in $\Lambda$CDM for different
degrees of non-Gaussianity. Solid, dotted and dashed lines show
results for the RGS $P(y)$ using $G=1$ (i.e. Gaussian), 2 and 5
respectively. The left-hand panel shows counts as a function of flux
at 143GHz, the centre panel shows the redshift distribution of
clusters brighter than $|S_{143{\rm GHz}}|=1$mJy, and the right hand
panel shows the distribution of angular sizes for the same
clusters. All results are for the parameters of the {\sc bolocam}
experiment and the $\Lambda$CDM cosmology.}
\label{fig:nongauss}
\end{figure*}

For the purposes of these examples, we consider only the three
cosmological parameters $\Omega_{\rm b}$, $\Omega_0$ and $\sigma_8$ to
be free (and assume $\Omega_0+\Lambda_0=1$). The effects of varying
$h$ may be simply scaled out of any results. These parameters are
already well-measured from analysis of CMB experiments
(e.g. \citealt{netter01}) and cluster abundances
(e.g. \citealt{ecf96}). We will therefore adopt prior probabilities
for the values of these parameters in our likelihood analysis. Each
parameter is assumed to have a Gaussian prior probability distribution
with a width as given in Table~\ref{tb:priors}, which also lists the
true value of the parameter assumed in our likelihood analysis (we
choose representative values, not necessarily the current best-fit
values, for this analysis). For $\Omega_0$ and $\Omega_{\rm b}$ we
adopt priors based on the analysis of CMB anisotropies of
\citep[][specifically their ``Flat, LSS \& SN1a''
determinations]{netter01}, while for $\sigma_8$ we use the fractional
errors obtain from cluster abundance analysis by \citet{ecf96}. We
will show results both with and without these priors. We will take a
model with $G=5$ as an example (note that \citet{rgs} were able to
find a model with $G=10$ and low $\sigma_8$ which produced the correct
$z=0$ cluster abundance and also that \citet{verde01} demonstrated
that stronger constraints could be placed on $G$ using the observed
cluster X-ray size-temperature relation).

\begin{table}
\caption{The priors used to obtain constraints on $G$ and
$K_{34}$. For each parameter we assume a prior probability which is a
Gaussian with $\sigma$ equal to the fraction given in the table.}
\label{tb:priors}
\begin{center}
\begin{tabular}{ccc}
Parameter & ``True'' value & Prior accuracy \\
\hline
$\Omega_0$ & 0.3 & 15\% \\
$\Omega_{\rm b}$ & 0.04 & 20\% \\
$\sigma_8$ & 0.9 & 8\% \\
\end{tabular}
\end{center}
\end{table}

Because of the prohibitively large amount of time required for a full
analysis of the likelihood contours for model parameters obtainable
from SZ surveys (even with only four parameters allowed to vary), we
will estimate likelihood contours using a Fisher-matrix analysis
(e.g. \citealt{jungman96}). This works well for the {\sc planck}
survey, where the Gaussian assumption made in the Fisher-matrix
analysis is justified. For the {\sc bolocam} survey however, this
assumption is rather poor. To demonstrate this we show in
Fig.~\ref{fig:consttest} the likelihood contours obtainable for a 1
square degree {\sc bolocam} survey of $|S_{143{\rm GHz}}|\geq1$mJy
clusters, if redshifts are measured for all clusters. The thin lines
indicate the results from the Fisher-matrix analysis (dashed lines
assume no priors, solid lines assume the priors from
Table~\ref{tb:priors}), marginalized over $\Omega_0$ and
$\sigma_8$. To compare to this, we generated a large, four-dimensional
grid of models and then generated a large number of Monte-Carlo
realizations of the observational dataset. The best fitting model to
the mock dataset was determined using a maximum likelihood approach,
and the results used to construct likelihood contours in the
$G$--$\Omega_{\rm b}$ plane. It is clear that the Gaussian assumption
used in the Fisher-matrix analysis is a poor approximation, with the
true likelihood contours depending on the parameters in a more complex
way. Nevertheless, the Fisher-matrix approach gives at least a rough
idea of the constraints that can be placed on model parameters (and is
very rapid to compute).

\begin{figure}
\psfig{file=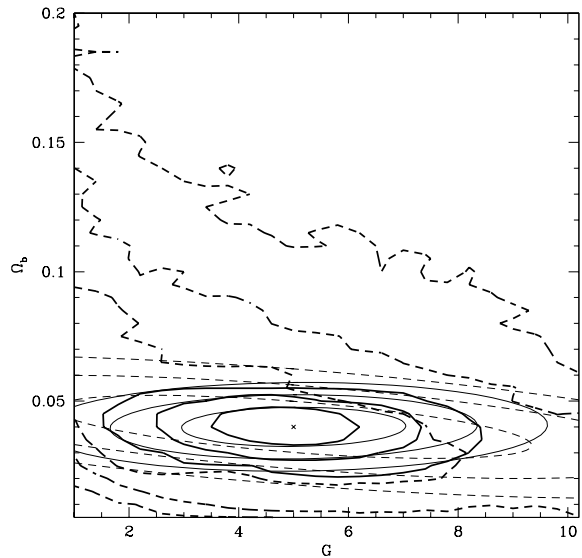,width=80mm}
\caption{Constraints in the $G$--$\Omega_{\rm b}$ plane obtainable
from the {\sc bolocam} survey assuming redshifts can be obtained for
all clusters brighter than $|S_{143{\rm GHz}}|=1$mJy. Thin lines are
the estimates from a Fisher matrix analysis, while thick lines were
obtained from a large number of Monte-Carlo realizations of the
dataset. Dashed lines show the results with no priors, while solid
lines show results using the priors of
Table~\protect\ref{tb:priors}. Contours show the 68.3\%, 95.4\% and
99.7\% confidence regions. The cross shows the true values of the
parameters.}
\label{fig:consttest}
\end{figure}

Figure~\ref{fig:constng} shows the constraints obtainable on the
non-Gaussianity parameter $G$ by the {\sc bolocam} and {\sc planck}
experiments. In each panel we marginalize over two of the free
parameters and show the error contours in the plane of the remaining
two parameters. Thick and thin lines show results with and without
priors (priors are included only for the {\sc bolocam} experiment ---
for the {\sc planck} experiment the priors of Table~\ref{tb:priors} do
not significantly alter the confidence regions). Dashed lines indicate
results when no redshift information is available, while solid lines
show results if redshifts are measured for all clusters.

\begin{figure*}
\hspace{5mm}\psfig{file=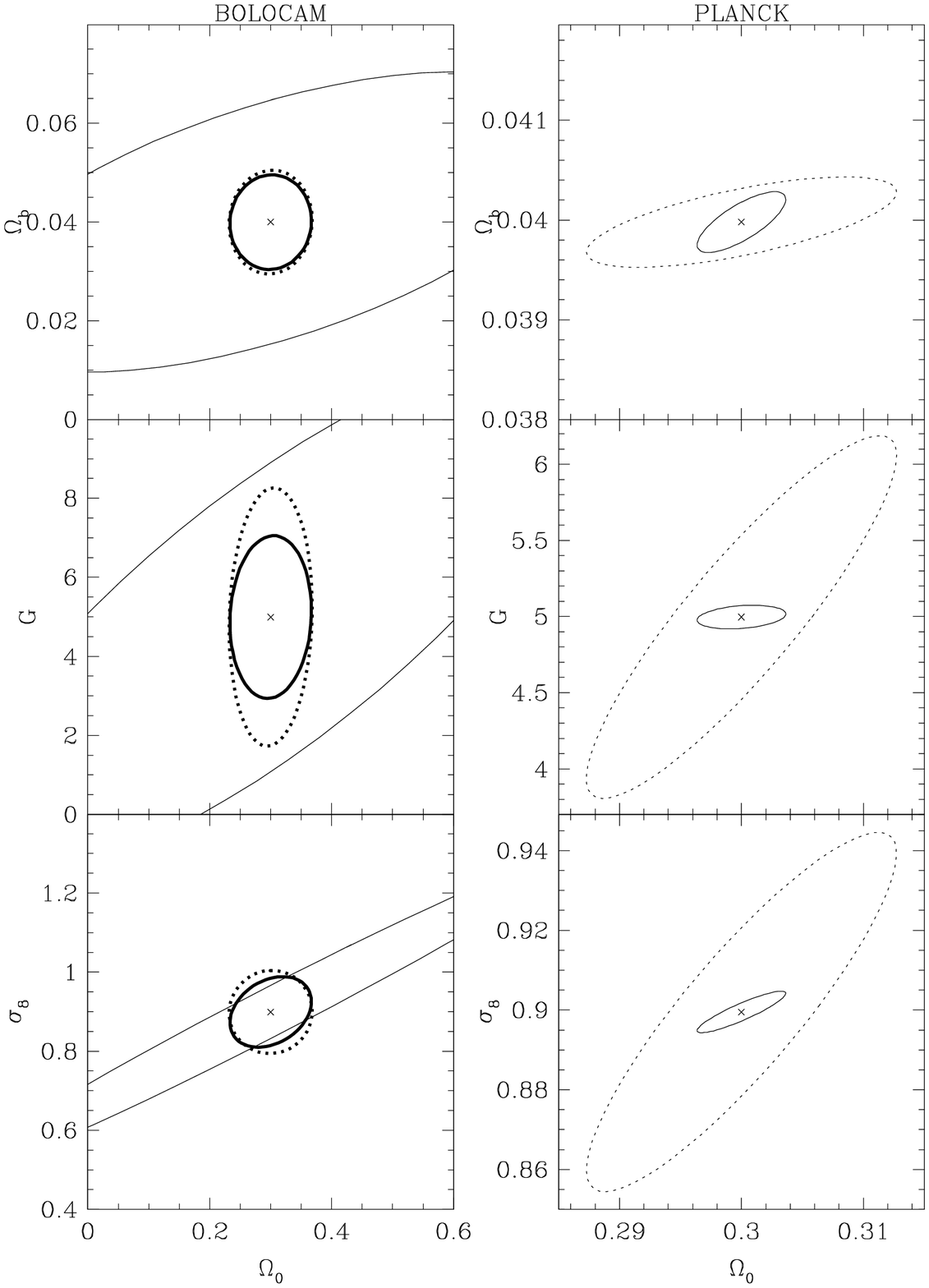,width=155mm,bbllx=5mm,bblly=20mm,bburx=188mm,bbury=265mm,clip=}
\caption{\emph{Left-hand panels:} Joint constraints on $\Omega_0$ and
$\Omega_{\rm b}$ (top panel), $\Omega_0$ and $G$ (middle panel) and
$\Omega_0$ and $\sigma_8$ (lower panel) obtainable from a 1 square
degree {\sc bolocam} survey to $|S_{143{\rm GHz}}|=1$mJy using a
Fisher matrix analysis. In each panel we have marginalized over the
two parameters not shown. Thin lines show results with no priors,
while thick lines use the priors from
Table~\protect\ref{tb:priors}. Dashed lines make use of only the
number counts of SZ clusters, while solid lines make use of the their
redshift distribution also (and assume that all clusters have measured
redshifts). The contours show the 68.3\% (i.e. $1\sigma$) confidence
region. \emph{Right-hand panels:} Same for a full-sky {\sc planck}
survey to $|S_{143{\rm GHz}}|=1$mJy. Note that we do not show lines
including priors for the {\sc planck} experiment, since these do not
significantly alter the confidence ellipses.}
\label{fig:constng}
\end{figure*}

\begin{figure*}
\begin{tabular}{c@{}c@{}c}
\psfig{file=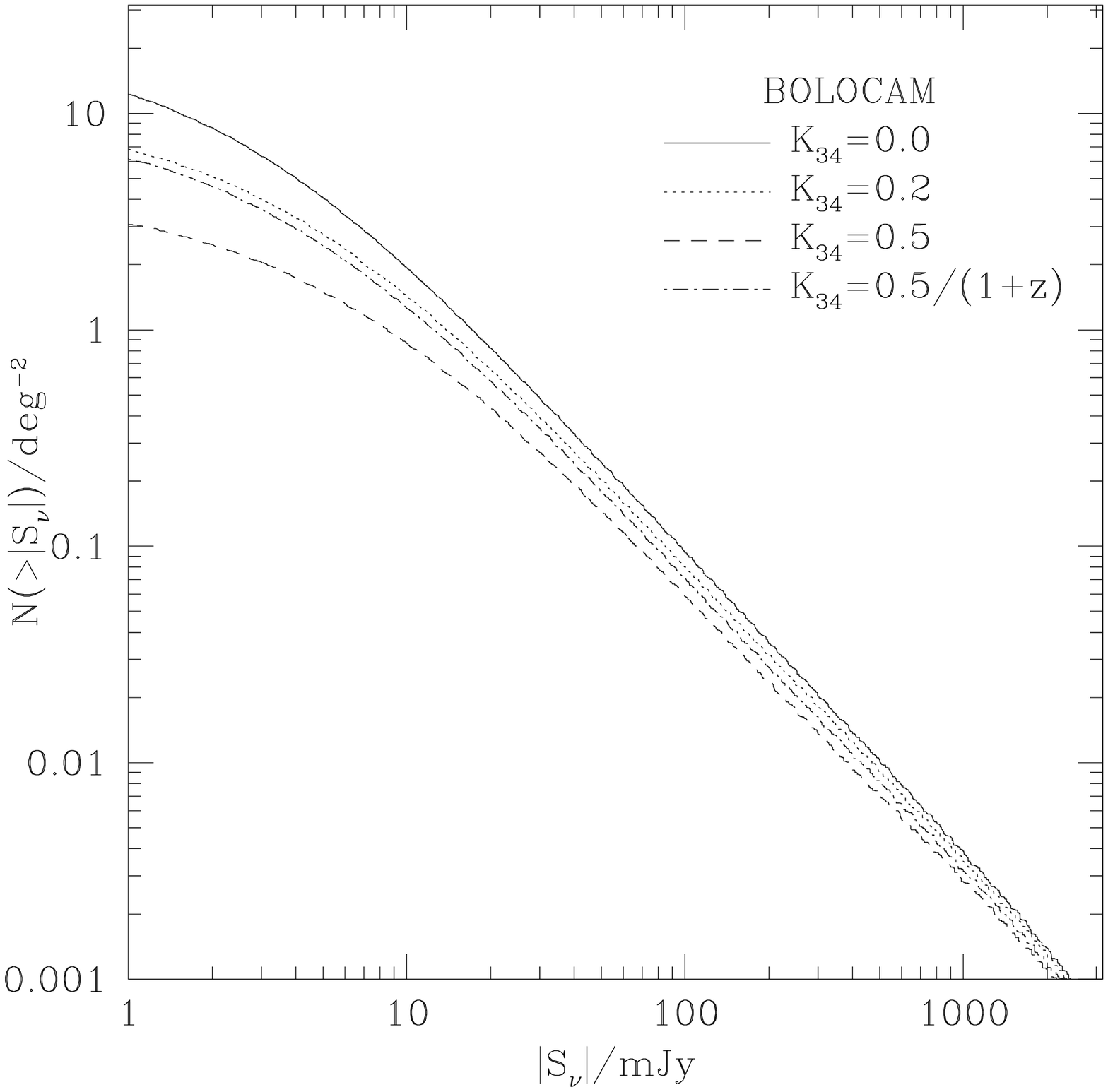,width=58mm} &
\psfig{file=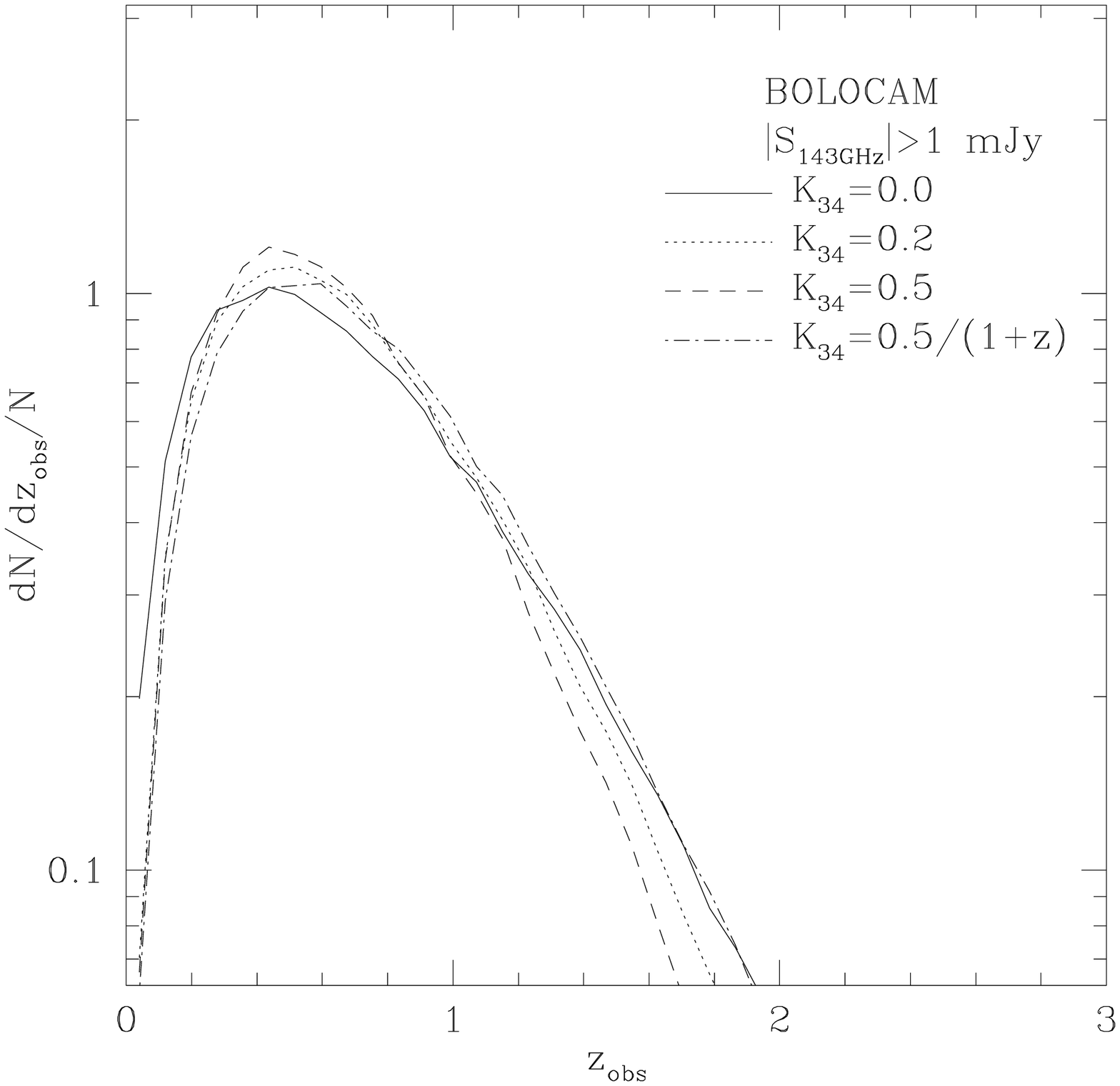,width=58mm} &
\psfig{file=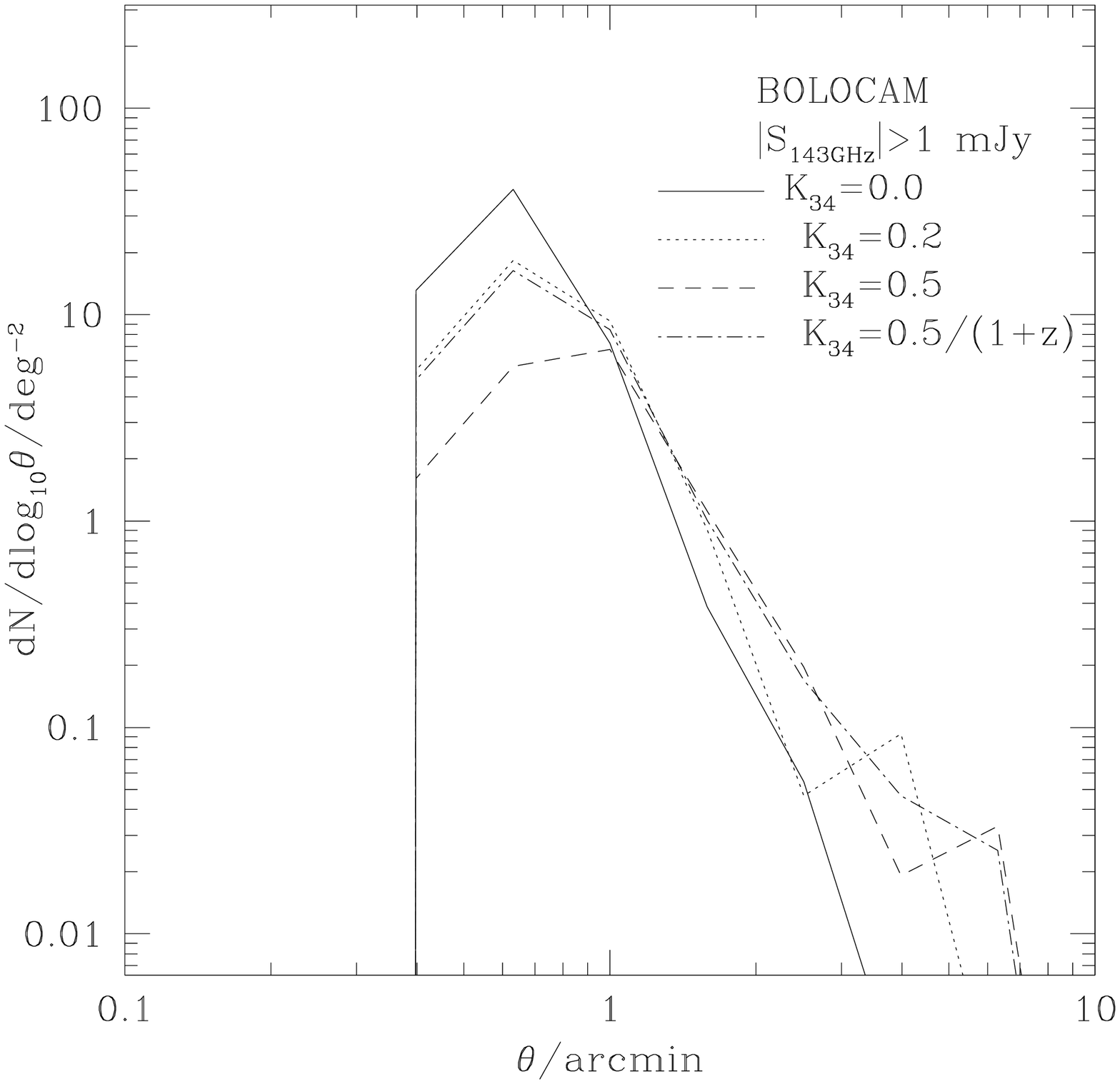,width=58mm}
\end{tabular}
\caption{Properties of SZ clusters in $\Lambda$CDM for different
levels of entropy injection. Solid, dotted and dashed lines show
results for $K_{34}=0.0$, 0.3 and 0.8 respectively. The dot-dashed
lines show the results when the entropy injected varies with the
formation redshift according to $K_{34}=0.8/(1+z_{\rm f})$. The
left-hand panel shows counts as a function of flux at 143GHz, the
centre panel shows the redshift distribution of clusters brighter than
$|S_{143{\rm GHz}}|=1$mJy, and the right hand panel shows the
distribution of angular sizes for the same clusters. All results are
for the parameters of the {\sc bolocam} experiment and the
$\Lambda$CDM cosmology with the J2000 $P(y)$.}
\label{fig:entropy}
\end{figure*}

It is immediately apparent that, for the {\sc bolocam} experiment, no
interesting constraints on $G$ can be obtained unless either priors
are assumed for the other parameters, or if redshifts are measured for
all clusters (preferably both). (Note that in the left-hand panels the
confidence ellipses for the case of no redshift information and no
priors is so large that they lie entirely outside the regions
plotted.)  With full redshift information and with the priors of
Table~\ref{tb:priors} we find that {\sc bolocam} will measure $G$ to
an accuracy of $\pm 40\%$ ($1\sigma$). (It must be kept in mind,
however, that the Fisher matrix analysis is only approximate for this
survey---see Fig.~\ref{fig:consttest}.) Interesting constraints on the
three cosmological parameters considered can also be obtained if
redshifts are available for all clusters in the {\sc bolocam} sample,
although there are strong degeneracies between the parameters
(e.g. between $\sigma_8$ and $\Omega_0$ in particular). The figure
also demonstrates degeneracies between the various parameters, in
particular between $\sigma_8$ and $G$. Increasing $\sigma_8$ or $G$
tends to increase the total number of clusters of all fluxes and also
to strongly increase the numbers seen at high redshifts (see
Figs.~\ref{fig:sigma8} and \ref{fig:nongauss}) leading to the
degeneracy seen in Fig.~\ref{fig:constng}.

For the {\sc planck} experiment simple number counts of SZ clusters
will constrain $G$ to an accuracy of $\pm 5\%$ while full redshift
information would reduce that uncertainty to approximately $\pm 2\%$
(since it is unlikely that redshifts would be available for all
clusters the actual constraint on $G$ will lie somewhere in between
these two values). Even without redshift information, the {\sc planck}
experiment will provide extremely tight constraints on cosmological
parameters through the SZ effect (which will supplement the
determinations from the primordial CMB anisotropies which are the main
goal of this experiment). It should be noted that, in the case of {\sc
planck} the tight constraints shown here do not account for the
systematic uncertainties due to theoretical uncertainties (see
\S\ref{sec:theorunc}), which would at present be the dominant source
of error.

\subsubsection{Constraints on Preheating}
\label{sec:resent}

Preheating of clusters will affect the total number counts (see
Fig.~\ref{fig:entropy} where we show results for a range of values for
the minimum entropy parameter $K_{34}$ defined in
\S\ref{sec:descpre}), by dramatically reducing the number of faint
clusters, while having very little effect on the redshift distribution
of SZ sources. Preheating also increases the sizes of observed
clusters, producing a tail to large $\theta$ in the distribution of
angular sizes. The effect is small and would require good statistics
(and so a significantly larger survey than the 1 square degree
proposed for {\sc bolocam}), it is a clear signal of entropy injection
since no other factor we examined has a similar effect on sizes.

Figure~\ref{fig:constpr} shows confidence regions for the entropy
parameter $K_{34}$ (the format follows Fig.~\ref{fig:constng}). Again
we see that useful constraints on $K_{34}$ from the {\sc bolocam}
experiment can be obtained only if redshift information is available
or if priors are assumed for the various cosmological
parameters. Assuming this to be the case, we find that {\sc bolocam}
can constrain $K_{34}$ to an accuracy of $\pm 60\%$. The vastly larger
area surveyed by {\sc planck} will allow it to constrain $K_{34}$ to
an accuracy of $\pm 6\%$ assuming full redshift information. Note
again that the effects on SZ statistics of altering $K_{34}$ are
degenerate with changes in all three of the cosmological parameters
considered in our analysis.

\begin{figure*}
\hspace{5mm}\psfig{file=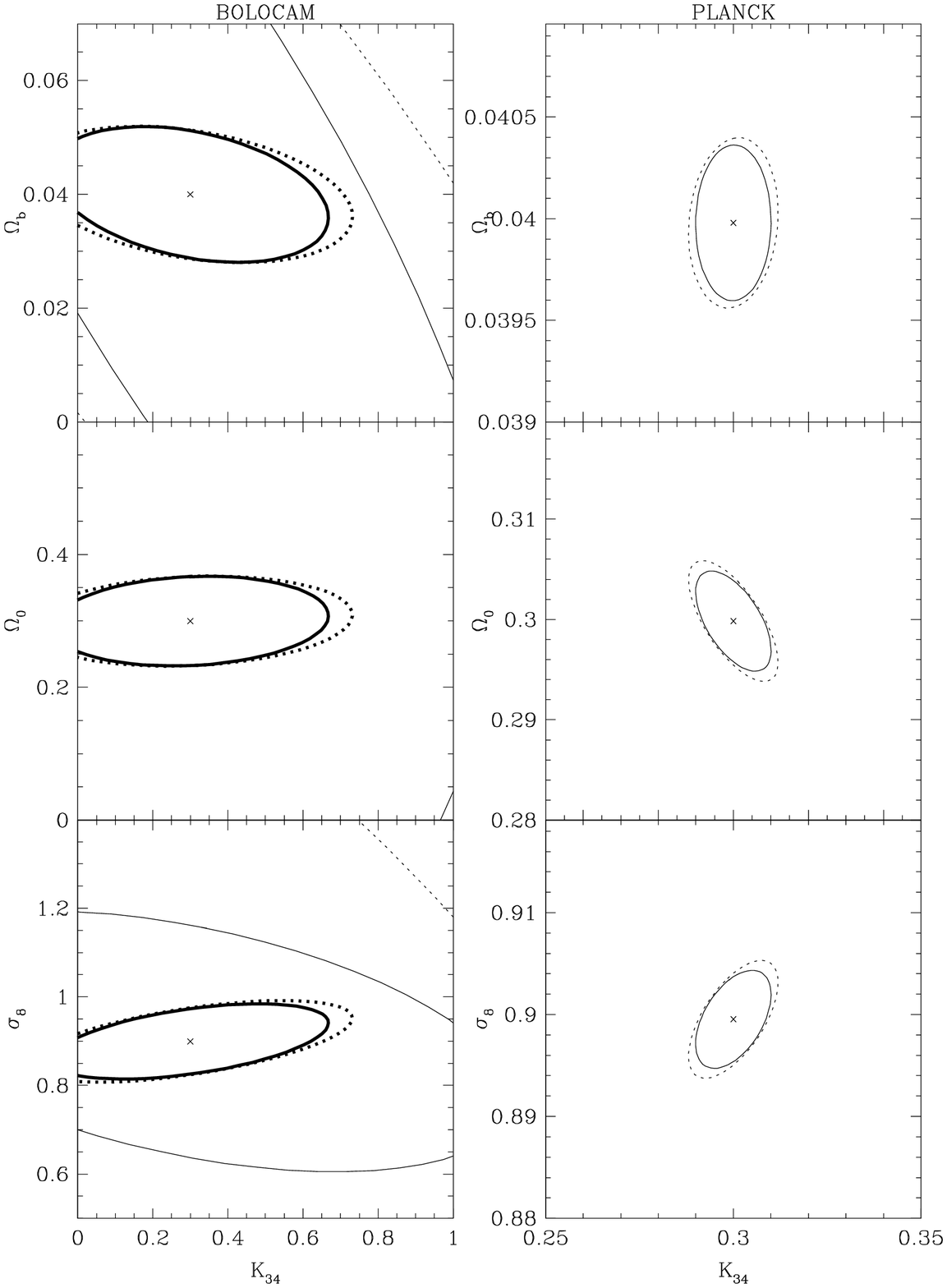,width=155mm,bbllx=5mm,bblly=20mm,bburx=190mm,bbury=265mm,clip=}
\caption{\emph{Left-hand panels:} Joint constraints on $K_{34}$ and
$\Omega_{\rm b}$ (top panel), $K_{34}$ and $\Omega_0$ (middle panel)
and $K_{34}$ and $\sigma_8$ (lower panel) obtainable from 1 square
degree {\sc bolocam} survey to $|S_{143{\rm GHz}}|=1$mJy using a
Fisher matrix analysis. In each panel we have marginalized over the
two parameters not shown. Thin lines show results with no priors,
while thick lines use the priors from
Table~\protect\ref{tb:priors}. Dashed lines make use of only the
number counts of SZ clusters, while solid lines make use of the their
redshift distribution also (and assume that all clusters have measured
redshifts). \emph{Right-hand panels:} Same for a full-sky {\sc planck}
survey to $|S_{143{\rm GHz}}|=1$mJy. Note that we do not show lines
including priors for the {\sc planck} experiment, since these do not
significantly alter the confidence ellipses.}
\label{fig:constpr}
\end{figure*}

\section{Discussion}
\label{sec:disc}

We have made an in-depth analysis of the statistical properties
(including counts, redshift and size distributions) of clusters that
should be detected through the thermal SZ effect in forthcoming
experiments, in particular {\sc bolocam} and {\sc planck}.

Our model is based upon simple analytic calculations allowing us to
explore a wide range of parameter space which is particularly
important given that the details of the gas distribution within
clusters are unknown, and also allows for rapid calculation of
confidence regions for cosmological and gas distribution
parameters. Our standard calculation consists of a distribution of
dark-matter halos with a mass function matched to that found in
numerical simulations and with a distribution of formation redshifts
motivated by the Press-Schechter theory. These halos are filled with
gas in hydrostatic equilibrium and the Compton $y$ parameter is
calculated as a function of cluster radius. Finally, we account for
smearing of the cluster SZ profile by the experimental beam and then
compute the observable SZ flux above the self-consistently determined
background. The result is the distribution of SZ clusters as a
function of total SZ flux, redshift and angular size.

We have quantified the effects of several theoretical uncertainties on
the statistical quantities measurable from SZ surveys. We find that
the uncertainty in how cluster formation redshifts determine the
thermodynamic properties of their gas leads to an uncertainty in
cluster number counts comparable to the statistical uncertainty
expected for the {\sc bolocam} survey. (Due to the broad redshift
distributions characteristic of the SZ effect we find that sample
variance is typically very small, i.e. close to the Poisson limit for
the 1 square degree {\sc bolocam} survey which we considered.) Other
uncertainties, such as the exact form of $P(y)$ and the density and
temperature profiles of cluster gas have slightly smaller effects, but
will still dominate over the statistical errors for a full-sky {\sc
planck} survey. As such, it will be crucial to further our
understanding of cluster gas properties before future SZ surveys can
be exploited fully.

Finally, we explored the constraints that may be set on cosmological
parameters and also the presence of any non-Gaussianity in the initial
conditions or preheating of ICM gas in the early Universe. Redshift
distributions are potentially very sensitive to $\Omega_0$, $\sigma_8$
and the presence of primordial non-Gaussianity. Preheating of clusters
has much less effect on the redshift distributions but can
significantly reduce the number of faint SZ clusters and also produces
a tail to large sizes in the SZ cluster size distribution. A
Fisher-matrix analysis reveals that {\sc bolocam} will provide
interesting constraints on cosmological parameters and the degree of
non-Gaussianity (modulo uncertainties in $K_{34}$) and/or preheating
only if redshifts are measured for all the detected clusters (or if
prior probability distributions are assumed for the cosmological
parameters). The {\sc planck} experiment, due to the large area it
will survey, should measure the non-Gaussianity or preheating
parameters, $G$ and $K_{34}$ respectively, to accuracies of 5--10\%
even without redshifts for any of the clusters. Furthermore, the {\sc
planck} SZ survey will provide independent and highly accurate
constraints on cosmological parameters that will complement those
obtained from analysis of primordial CMB anisotropies.

In conclusion, surveys in the SZ effect will soon provide valuable
measurements of various cosmological parameters which will compliment
measurements from other techniques. Assuming that the theoretical
uncertainties highlighted in this paper can be adequately resolved,
the SZ effect also has the potential to constrain the presence of
non-Gaussianity and also to provide strong constraints on the
distribution of gas within clusters and the presence or otherwise of
any preheating.

\section*{Acknowledgements}

AJB acknowledges helpful conversations with Scott Kay and Peng Oh. The
Hubble Volume numerical simulations were kindly made available by the
VIRGO Consortium. This work was supported in part by NSF AST-0096023,
NASA NAG5-8506, and DoE DE-FG03-92-ER40701.

\end{document}